%% file: main.tex
\documentclass[journal]{IEEEtran}
\usepackage[utf8]{inputenc} 
\usepackage[T1]{fontenc}
\usepackage{url}
\usepackage{ifthen}
\usepackage{balance}

\usepackage[fleqn]{nccmath} 
\usepackage{cite}             
\usepackage{dsfont}
\usepackage{soul} 
\usepackage{amssymb} 
\usepackage{amsmath}
\usepackage{amsthm} 
\usepackage{aligned-overset}
\usepackage[bookmarks=false]{hyperref}
\newtheoremstyle{mytheorem}
  {0pt}
  {0pt}
  {\itshape}
  {}
  {\bfseries}
  {:}
  {0.5em}
  {}

\theoremstyle{mytheorem}
\newtheorem{theorem}{Theorem}
\newtheorem{lemma}{Lemma}

\newtheorem{remark}{Remark}
\newtheorem{definition}{Definition}

\newcommand{\floor}[1]{\left \lfloor {#1} \right \rfloor }

\newcommand{\norm}[1]{\left\Vert#1\right\Vert}

\newcommand{\n}[1]{{\|\mathbf{#1}\|}} 

\newcommand{\bx}{\mathbf{x}}
\newcommand{\cV}{{\cal V}}
\newcommand{\cC}{{\cal C}}
\newcommand{\cS}{{\cal S}}
\newcommand{\R}{\mathbb{R}}
\def\b{\mathbf}
\NewDocumentCommand\sqn{mg}{%
    \|\mathbf{#1}_{\IfNoValueTF{#2}{}{#2}}\|^2%
}

\DeclareMathOperator{\EX}{\mathbb{E}}

\usepackage{graphicx}
\graphicspath{{figs/}}
\usepackage{subcaption}

\usepackage{tikz}
\usepackage{pdfpages} 
\usetikzlibrary{math} 

\usetikzlibrary{decorations.pathreplacing,calligraphy}
\usepackage{subfiles} 
\begin{document}

\title{\textcolor{black}{On PIR and SPIR Over Gaussian MAC}}


\author{%
 \IEEEauthorblockN{Or Elimelech and Asaf Cohen}
 \IEEEauthorblockA{\\The School of Electrical
and Computer Engineering\\
                   Ben-Gurion University of the Negev, Israel\\}
}
\maketitle
\input{abstract}
\input{Intorduction}
\input{ProblemStatement}

\input{Joint_FadingNew}

\input{ComparisonFadingSchemes}
\input{SPIR}

\input{SPIR_No_Key}
\input{Conclusion}

\input{appendices}

\balance

\bibliographystyle{IEEEtran}
\bibliography{references}


\begin{IEEEbiography}[{\includegraphics[width=1in,height=1.25in,clip,keepaspectratio]{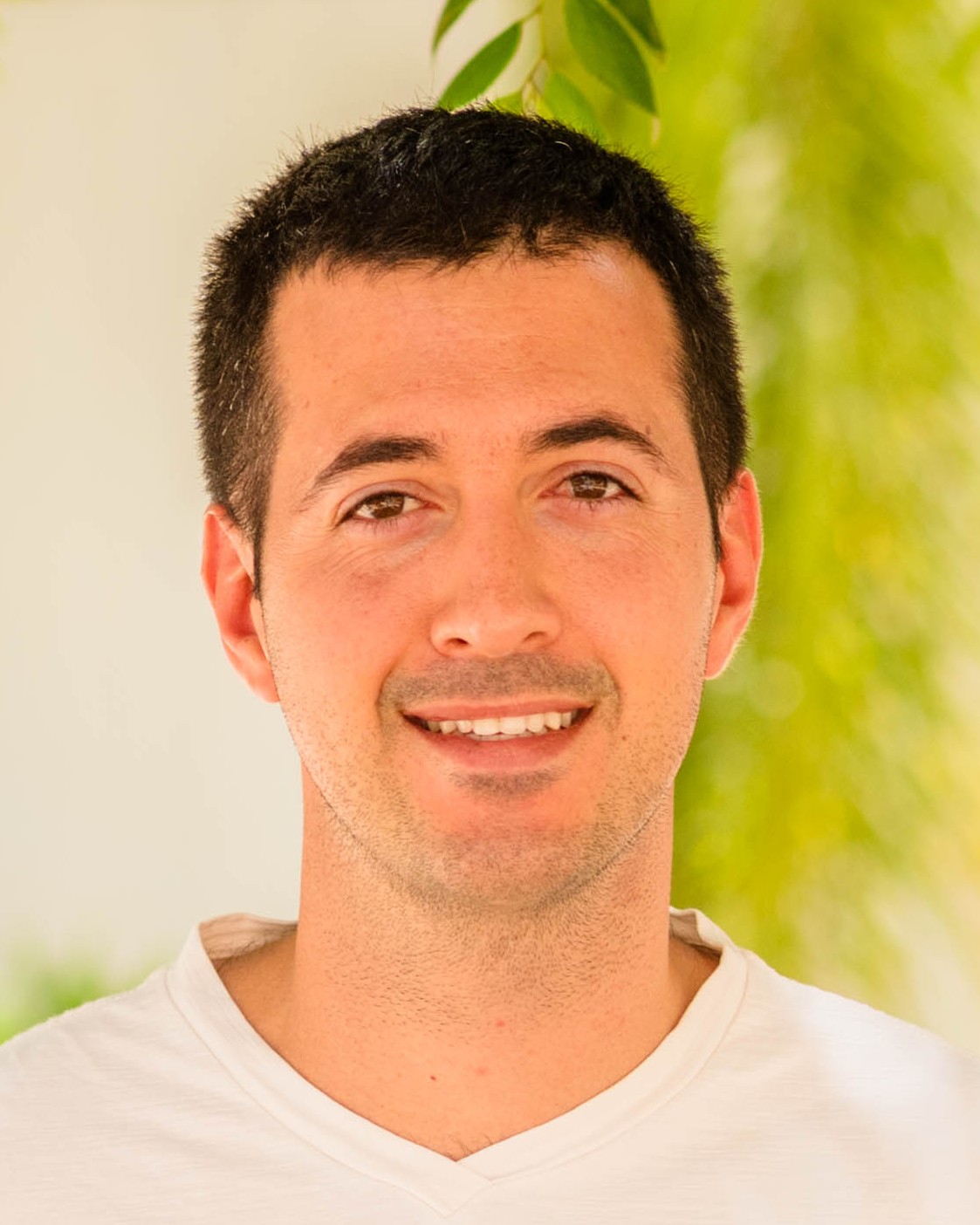}}]{Or Elimelech}
received his B.Sc. degree in Electrical Engineering and his M.Sc. degree in Communication Systems Engineering from Ben-Gurion University of the Negev, Israel, in 2021 and 2023, respectively. He is currently pursuing a Ph.D. in the School of Electrical and Computer Engineering at the same institution. 
His areas of interest include information theory and physical layer security, with recent works exploring privacy and secrecy in modern communication networks, as well as semantic communication.
\end{IEEEbiography}

\begin{IEEEbiography}[{\includegraphics[width=1in,height=1.25in,clip,keepaspectratio]{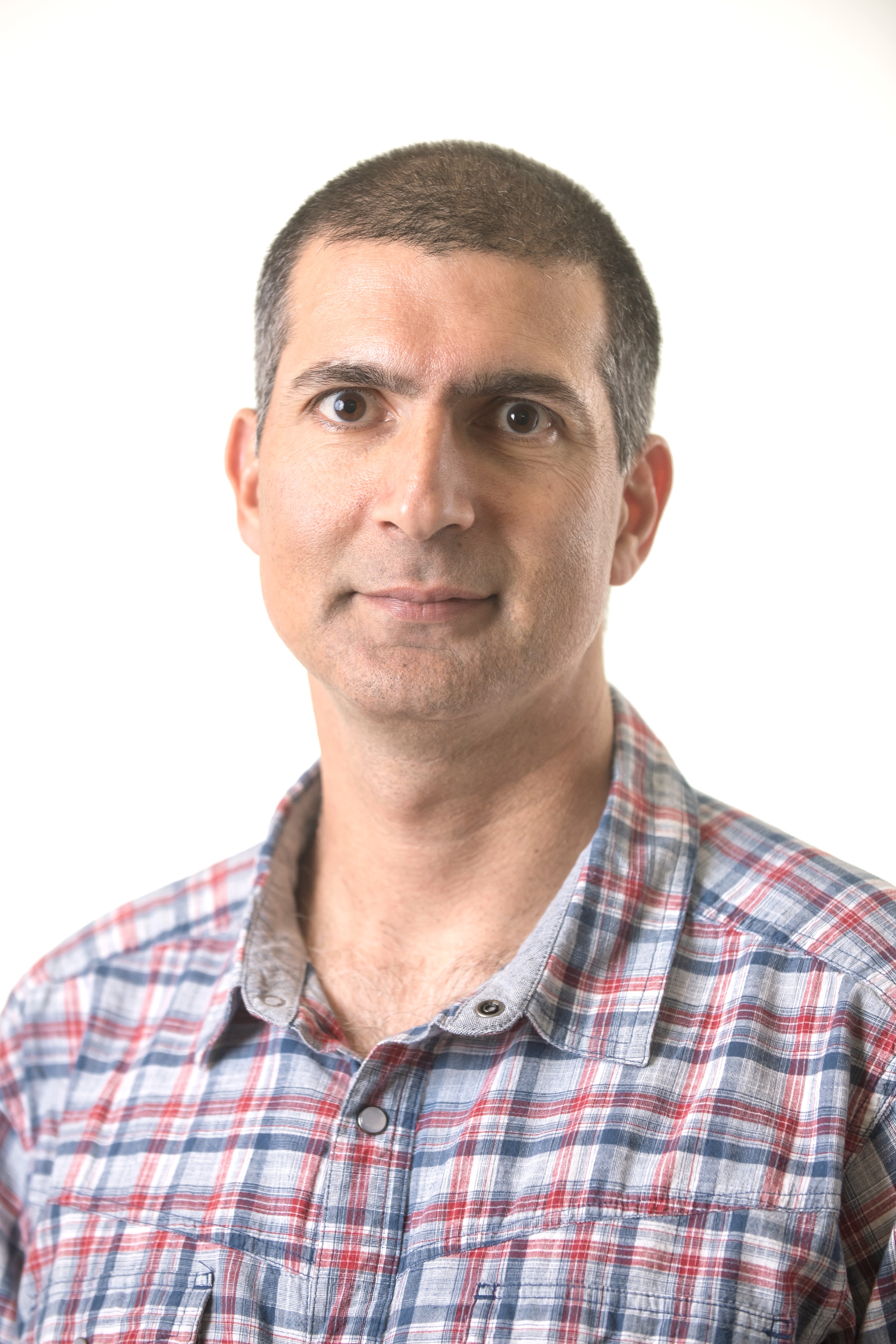}}]{Asaf Cohen}
received the B.Sc. (Hons.), M.Sc. (Hons.), and Ph.D. degrees from the Department of Electrical Engineering, Technion, Israel Institute of Technology, in 2001, 2003, and 2007, respectively.
From 1998 to 2000, he was with the IBM Research Laboratory, Haifa, where he was working on distributed computing.
Between 2007 and 2009 he was a Post-Doctoral Scholar at the California Institute of Technology, and between 2015–2016 he was a visiting scientist at the Massachusetts Institute of Technology.
He is currently an Associate Professor and the Vice Chair for Teaching at the School of Electrical and Computer Engineering, Ben-Gurion University of the Negev, Israel.
His areas of interest are information theory, learning, and coding. In particular, he is interested in network information theory, network coding and coding in general, network security and anomaly detection, statistical signal processing with applications to detection and estimation and sequential decision-making. He received several honors and awards, including the Viterbi Post-Doctoral Scholarship, the Dr. Philip Marlin Prize for Computer Engineering in 2000, the Student Paper Award from IEEE Israel in 2006 and the Ben-Gurion University Excellence in Teaching award in 2014. He served as a Technical Program Committee for ISIT, ITW and VTC for several years, and as an Associate Editor for Network Information Theory and Network Coding; Physical Layer Security; Source/Channel Coding and Cross-Layer Design to the IEEE Transactions on Communications.
\end{IEEEbiography}


\end{document}

%% file: abstract.tex
\begin{abstract}
\textcolor{black}{This paper} revisits the problems of Private Information Retrieval (PIR) and Symmetric PIR (SPIR).
\textcolor{black}{In PIR, a user retrieves a desired message from $N$ replicated, non-communicating databases, each storing the same $M$ messages, while preserving the privacy of the requested message index.}
SPIR extends this notion further by additionally protecting the privacy of the databases, ensuring that the user learns no information beyond the requested message.

\textcolor{black}{In this paper,} we assume a block-fading Additive White Gaussian Noise Multiple Access Channel (AWGN MAC) linking the user and the databases.

Previous work by Shmuel et al. presented a joint channel-PIR scheme utilizing the Compute and Forward (C\&F) protocol, demonstrating the potential of a joint PIR-channel coding scheme over a separated one, yet still lagging behind the channel capacity and requiring significant computational complexity. 
We propose an improved scheme that offers reduced computational complexity while improving the achievable rate for finite parameters, as well as its scaling laws.
Specifically, the achievable rate outperforms the C\&F-based approach and scales with the number of databases $N$ and the power $P$ similarly to the channel capacity \textit{without the privacy constraint}. 
Furthermore, the analysis demonstrates that the improved rate exhibits only a finite gap from this unconstrained channel capacity -- $1$ $bit/sec/Hz$ as $N$ increases. 

\textcolor{black}{Finally, we provide two SPIR schemes.
The first is a modification for our PIR scheme to attain SPIR with no rate
loss, which is accomplished by introducing shared common randomness among databases. 
The second is a novel joint channel-SPIR scheme that utilizes the channel and lattice codes characteristics to nontrivially achieve SPIR without requiring common randomness,
at the price of a loss in the achievable rate.}
\end{abstract}

%% file: Intorduction.tex
\section{Introduction}
Private Information Retrieval (PIR) deals with a user wishing to retrieve a message from a database while keeping the identity of the desired message secret from the database itself.

PIR has been extensively studied within the Computer Science community, with significant contributions from \cite{gasarch2004survey, ostrovsky2007survey, yekhanin2010private}. These studies primarily focused on computational solutions, leading to Computational PIR (CPIR). 
The information theory community has recently started exploring PIR, offering a unique interpretation to understand the problem's fundamental limits.
The information-theoretic approach to PIR seeks to achieve \textit{perfect information privacy}, that is, the identity of the desired message remains concealed even against unlimited computational power.
Furthermore, an information-theoretic perspective typically assumes that messages are large enough, thereby neglecting the cost of communicating the queries themselves.

In the classical setup, there are $N$ identical databases (or servers), each containing the same $M$ messages.
These databases do not communicate with each other. 
A user, who wishes to retrieve a specific message without revealing its index to the databases, formulates a series of queries.
The goal is to reduce the overhead necessary for maintaining privacy.

The pioneering work presented in \cite{chor1995private} explored whether employing multiple databases could lead to a more efficient PIR solution. 
\textcolor{black}{It was shown} that in a single-database scenario, perfect information-theoretic privacy can only be attained by downloading the entire database.
This means that for $M$ messages, the PIR rate is $1/M$.
They further established that when using only two databases ($N=2$), the rate can be improved to $1/2$, irrespective of the number of messages $M$.

Numerous PIR schemes have drawn inspiration from the two-database scheme proposed in \cite{chor1995private}. 
The well known PIR capacity given by ${C_{PIR}=(1-1/N)(1-(1/N)^M)}$ was established in \cite{sun2017capacity}.
Over the years, many extensions to the PIR problem have been explored.
For example, robust PIR where some of the databases may fail to respond and $T$-private PIR where even if any $T$ of the $N$ databases collude, the identity of the retrieved message remains completely unknown to them \cite{sun2017capacityColluding}.
PIR from byzantine databases, where any $B$ databases could provide erroneous responses intentionally or unintentionally \cite{banawan2018capacityByzantine}.
Cache-aided PIR considers another case when the user caches part of the messages in advance, and the idea is to exploit cached data as side information\cite{tandon2017capacity,wei2018cache,wei2018fundamental,wei2019capacity,seo2021fundamental,vaidya2023cache,vaidya2024multi}.

All works mentioned above consider uncoded storage, that is, the messages are replicated on the databases. 
Although it is reliable against database failure, it is not necessarily efficient in terms of storage capacity.
Interesting extensions considering PIR using coded databases can be found in \cite{Banawan2016TheCO,wang2017symmetric,tajeddine2017private,chan2015private,tajeddine2018private,zhu2019new,zhou2020capacity,tajeddine2019private,li2020towards}.

\textcolor{black}{Another important extension is Symmetric PIR (SPIR), first introduced in \cite{gertner1998protecting}. 
SPIR imposes stricter privacy requirements: It does not only protect user's privacy, but also ensures the security of the database by preventing the user from gaining any information beyond the requested message. 
It was demonstrated in \cite{gertner1998protecting} that SPIR cannot be achieved over orthogonal channels without shared common randomness among the databases. The capacity of SPIR in the classical setting was fully characterized in \cite{sun2018capacity}.}

The relevance of the PIR problem has grown in recent years, driven by the increasing need for privacy-preserving information retrieval in various domains. 
For example, in machine learning, PIR has gained attention as a means to securely access distributed datasets while maintaining data privacy \cite{wang2023fully,vithana2022efficient,kim2022information}.

While the PIR problem has been extensively studied over the past decades, most research has focused on simple communication channels characterized by \emph{orthogonal and noiseless links between the user and the database}. 
PIR over Noisy and orthogonal channels (NPIR) was examined in \cite{banawan2019noisy}, \textcolor{black}{showing} that the channel coding needed for combating channel errors is “almost separable" from the retrieval scheme and \textcolor{black}{requires only to agree on the traffic ratio.} 
Moreover, \cite{banawan2019noisy} considered PIR over different types of MAC and showed that, unlike NPIR, the channel coding and retrieval schemes cannot be separated in general.
Inspired by the work in \cite{banawan2019noisy}, the authors in \cite{shmuel2021private}\footnote{In this work, we refer to the results from \cite{shmuel2021private}. Please note that a correction to this paper has been made, as detailed in \cite{ElimelechCorre2024}.} considered PIR over a Gaussian MAC, i.e., where user-database communication takes place via a block-fading Gaussian MAC.
This model reflects a more realistic scenario for contemporary communication networks, which are dominated by wireless channels that require handling noise and interference.
In such a model, the scheme design has to consider noise induced by the channel and the ability to decode a mixture of the answers that share the medium while retaining privacy.
\textcolor{black}{It has been shown in \cite{shmuel2021private} that separating the channel code from the private retrieval code is sub-optimal; that is, applying an optimal channel code followed by an optimal PIR code does not achieve the best performance.
Moreover, the authors proposed a \textit{joint} PIR and channel coding scheme that leveraged the linearity of the channel while utilizing the C\&F coding scheme \cite{nazer2011compute} and modulo-lattice additive noise (MLAN) channel techniques \cite{erez2004achieving}.
Their approach demonstrated superior performance compared to a separation-based scheme, both with and without fading.}
However, the achievable rate in \cite{shmuel2021private} does not scale optimally with the power parameter $P$ (this is discussed in detail in Section \ref{sec:compare}).
Furthermore, the scheme proposed is computationally complex: while the C\&F  allows the receiver to decode a suitably chosen integer linear combination of the transmitted messages, the integer coefficients should be adapted to the channel fading state. 
Optimizing these coefficients is a Shortest Lattice Vector (SLV) problem.
It has been shown in \cite{sahraei2014compute} that this problem can be solved with polynomial complexity, that is, $O\left(N^2\sqrt{1+P\|\b{h}\|^2}\right)$, yet it is still complicated for large $N$ as it should be solved for each block.

\textcolor{black}{The main contributions of this work are as follows:} 
\begin{itemize}
    \item \textcolor{black}{ A novel PIR scheme over a block-fading AWGN MAC that narrows the gap from capacity, staying only $1$ $bit/sec/Hz$ from the unconstrained capacity when $N$ approaches infinity.
    It is critical to note that this result, formally stated in Lemma \ref{lemma:Gap} on the revised version, does not hold for the previous PIR scheme presented in \cite{shmuel2021private} for the block fading case.}
    \item \textcolor{black}{The scheme simplifies the computational complexity 
    by eliminating the need to solve the SLV problem required by the C\&F-based scheme in \cite{shmuel2021private}, which is known to be NP-hard in general.
    In contrast, our approach relies on straightforward linear operations combined with lattice quantization, avoiding combinatorial search and significantly reducing computational complexity.}
    \textcolor{black}{Although the asymptotic results hold for $N$ approaching infinity, numerical evaluations demonstrate that the proposed scheme achieves near-optimal performance with a relatively small number of databases, making it relevant for practical scenarios.}
    \item   \textcolor{black}{We extend our PIR scheme to the SPIR problem, ensuring user and database privacy.
    Our SPIR scheme achieves the same rate as the PIR scheme, provided the databases share a common random variable.}    
    \item   \textcolor{black}{ We further introduce a second SPIR scheme that does not require common randomness at all. 
    Unlike SPIR with orthogonal channels and non-communicating databases, where it has been shown that a SPIR scheme cannot be built without sharing some common randomness among the databases \cite[claim 3]{gertner1998protecting} (a condition challenging to implement in practice), our new scheme leverages the inherent characteristics of the AWGN MAC and a different lattice code construction to achieve SPIR, simplifying practical deployment significantly. 
    However, this advantage comes at the cost of a reduced retrieval rate.}
\end{itemize}

A preliminary version of this work was presented at the 2024 IEEE International Symposium on Information Theory (ISIT) \cite{10619535}. 
This paper distinguishes itself by offering full proofs and additional simulation results.
\textcolor{black}{
In addition, we provide two SPIR schemes: one that requires shared randomness among the servers and a novel one that does not require such shared common randomness.} 

%% file: ProblemStatement.tex
\section{System Model and Problem Statement}\label{Sec-Model and Problem Statement}

\subsection{Notational Conventions}
Throughout the paper, we will use boldface lowercase to refer to vectors, e.g., $\b{h} \in \R^L$, and boldface uppercase to refer to matrices, e.g., $\b{H} \in \R^{M \times L}$.  For a vector $\b{h}$, we write $\n{h}$ for its Euclidean norm, i.e. $\n{h} \triangleq \sqrt{\sum_{i}h_i^2}$. We denote by $\b{e}_i$ the unit vector with $1$ at the $i$th entry and zero elsewhere. We assume that the $\log$ operation is with respect to base 2.

\subsection{System Model}\label{sec-system model}
Consider the PIR problem in a basic setting with $N$ non-communicating databases. 
Each database stores the same set of messages $W_1^M=\{W_1,W_2,...,W_M\}$ , where $W_m$ is an $L$-length vector picked uniformly from prime-size finite field $\mathbb{F}_p^L$, where $p$ is a prime number.
These messages are independent and identically distributed, i.e., \textcolor{black}{
\begin{equation}\label{equ-Entropy of messages}
\begin{aligned}
H(W_1^M)=\sum_{l=1}^M H(W_l)= ML\log{p}.
\end{aligned}
\end{equation}
}
In PIR, the user wishes to retrieve the message $W_\theta$, assuming $\theta$ is  uniformly distributed over $[1,...,M]$, 
while keeping the index $\theta$ secret from each database.
To achieve private retrieval, the user generates a set of $N$ queries $Q_1(i), Q_2(i), ..., Q_N(i)$, where $i$ denotes a specific realization of $\theta$, one for each database, and each query is statistically independent of the messages, i.e.,
\begin{equation}\label{eq:const_querry}
I(W_1^M;Q_1(\theta),...,Q_N(\theta))=0.
\end{equation}
The $k$th database responds to its query $Q_k(i)$ with a message (or codeword) $\mathbf{x}_k(i)$ of fixed size $n$.
We follow the usual Gaussian MAC setup in the literature \cite{cover2012elements}, where a codeword is transmitted during $n$ channel uses.
The response $\mathbf{x}_k(i)$ is a deterministic function of the messages and the query. 
Therefore, for each $k\in\{1,...,N\}$, we have:
\begin{equation}\label{eq:const_deter}
H(\b{x}_k(\theta) | W_1^M,Q_k(\theta))=0.
\end{equation}
To ensure privacy, the query should not reveal the desired index $i$ to the database. 
Thus, we impose the privacy constraint, that is, for each database $j$, the random variable $\theta$ is independent of the query, the answer, and the messages:
\begin{equation}\label{eq:User-privacy}
\begin{split}
&\hspace{-0.7cm}\text{\textit{[User-privacy]}}\\
&I(\theta;Q_j(\theta),\mathbf{x}_j(\theta),W_1^M)=0 \ \text{for all } j\in\{1,...,N\}.
\end{split}
\end{equation}

\noindent \textcolor{black}{Later, in section \ref{sec:symm}, we modify the user-privacy condition in \eqref{eq:User-privacy} to appropriately address the requirements of the SPIR problem.}

Next, let us define the SPIR extension, which also requires database privacy (DB-Privacy).
That is, we want the user to gain no knowledge about any undesired messages. 
In our context, this is mathematically expressed as: \begin{equation}\label{eq:DB-privacy}
\begin{split}
    &\text{\textit{[DB-privacy]}}\\
    &I(W_{\overline{i}};Q_{1:N}(\textcolor{black}{i}),f(\bx_1(i),\dots,\bx_k(i)),\mathcal{K}) = 0 \quad \forall i\in\{1,...,M\},
\end{split}
\end{equation}
 where $W_{\overline{i}}=\left(W_1,\dots,W_{i-1},W_{i+1},\dots,W_M\right)$,
 $f(\bx_1(i),\dots,\bx_k(i))$ denotes the MAC output function.
 For example, in \eqref{eq:MAC_output_model} below, it is $\mathbf{y}$.
 $\mathcal{K}$ denotes the set of parameters, assumed to be available for the user, \textcolor{black}{such as channel state information and parameters related to the scheme.}

The databases are linked to the user via a block-memoryless fading AWGN channel (Figure \ref{fig:PIR_AWGN_MODEL}).
In this setup, the channel remains constant throughout the transmission of codewords of size $n$, and each block is independent of the others.
Thus, over a transmission of $n$ symbols, the user observes a noisy linear combination of the transmitted signals,
\begin{equation}\label{eq:MAC_output_model}
\mathbf{y}=\sum_{k=1}^{N}h_k\mathbf{x_k}+\mathbf{z}.
\end{equation}
Here, $h_k\thicksim\mathcal{N}(0,1)$ represents the real channel coefficients, and  $\mathbf{z}$ is an i.i.d. Gaussian noise $\mathbf{z}\thicksim\mathcal{N}(0,\mathbf{I}^{n\times n})$.
Additionally, we assume a per-database power constraint, where all transmitting databases operate with a fixed power $P$, and power cannot be allocated differently to different databases.
Thus, the transmitted codebook $\cC$ must satisfy the average power constraint, i.e., $\EX\left[\|\b{x}_k\|^2\right] \leq nP$.

\subfile{figs/PIR_AWGN_MAC_Model}

We assume there is channel state information at the transmitter (CSIT), i.e., $\{h_k\}_{k=1}^N$ are known at the transmitter.
Upon receiving the mixed response $\mathbf{y}$ from all the databases, the user decodes the required message $W_i$. 
Let $\widehat{W_i}$ denote the decoded message at the user and define the error probability for decoding a message as follows,
\begin{equation}\label{equ-probability of error definition}
P_e(L)\triangleq P_r(\widehat{W_i} \neq W_i).
\end{equation}
We require that $P_e(L)\rightarrow 0$ as $L$ tends to infinity.



\subsection{Performance Metric}\label{sec-performance Metric}
The PIR rate is typically defined as the ratio between the number of desired bits and the total number of received bits. 
In the information-theoretic formulation, where the size of the messages is assumed to be arbitrarily large, the upload cost is negligible.
Hence, the PIR rate can be expressed as:
\begin{equation}\label{equ-original PIR rate}
R_{PIR}\triangleq \frac{L\log{p}}{D},
\end{equation}
where $D$ is the total number of bits that have been downloaded.
However, counting the total number of downloaded bits is, by definition, suitable for cases where there is a separation between the PIR code and the channel code, or, in other words, the channels over which the PIR-coded data is sent are clean and orthogonal.
Thus, we define the PIR capacity over an AWGN MAC as follows.
\begin{definition} Denote:
 $$R_{PIR}^{MAC}(n)\triangleq \frac{H(W_i)}{n}=\frac{L\log{p}}{n},$$ 
 where $n$ represents the number of channel uses.
 The PIR capacity over AWGN MAC, denoted by $C_{PIR}^{MAC}$, is the supremum of $R_{PIR}^{MAC}(n)$, under which reliable communication is achievable, ensuring the privacy of the user's will. 
 That is, satisfying (\ref{eq:const_querry},\ref{eq:const_deter},\ref{eq:User-privacy}) and (\ref{equ-probability of error definition}). 
\end{definition}
\textcolor{black}{
Neglecting the privacy constraint simplifies the problem to an AWGN-MAC with a per-antenna power constraint. 
It is important to note that the MISO channel model is not applicable in this scenario, as the databases cannot cooperate.
This subtle distinction is crucial for understanding the problem's fundamental limits.
The sum-rate capacity for the AWGN-MAC under these conditions, assuming globally known and fixed channel coefficients, is given by
\cite{cover2012elements},
\begin{equation}\label{equ-MISO sum capacity with per-antenna power constraint}
C_{SR}^{MAC}=\frac{1}{2}\log\left( 1+ P\left(\sum_{k=1}^N |h_k|\right)^2\right).
\end{equation}
Hence, it becomes natural to use Gaussian MAC sum-rate capacity as an upper bound:  $C_{SR}^{MAC}\geq C_{PIR}^{MAC}$.
In fact, we will see that comparing the PIR rate to the channel capacity without the privacy constraint is quite an interesting comparison, as $C_{PIR}^{MAC}$ will approach it in several cases.
\textcolor{black}{It has already been shown that separating the channel coding from the PIR coding is not always optimal when dealing with a MAC (\!\!\cite{banawan2019noisy,shmuel2021private}).} 
Indeed, \cite{shmuel2021private} showed that better overall performance can be achieved when the PIR and the channel coding are jointly designed.
}

The subsequent section outlines our main results.
Section \ref{sec:Achievability} presents the achievability proof.
The achievable rate is compared to the result in \cite{shmuel2021private} and to the full channel capacity without any privacy constraint in Section \ref{sec:compare}.
In section \ref{sec:symm}, we discuss the SPIR problem and provide two SPIR schemes.


\textcolor{black}{
\subsection{Lattices \& Nested Lattice Codes}
}
\textcolor{black}{
Lattice codes are known by their ability to achieve the full capacity of the point-to-point AWGN channel \cite{erez2004achieving,ling2014achieving,liu2018construction,campello2018awgn}. 
We now provide a brief background on lattice codes, which will be useful in the remainder of this paper.
}

\textcolor{black}{
An $n$-dimensional lattice $\Lambda=\left\{\lambda=G\cdot\mathbf{i} \text{ : } \mathbf{i} \in \mathbb{Z}^n \right\}$ is a discrete subgroup of the Euclidean space $\mathbb{R}^n$  where $G\in \mathbb{R}^{n\times n}$ is called the generator matrix and its columns are linearly independent.
The lattice is closed under reflection and real addition.}

\textcolor{black}{
\begin{definition}[Quantizer]
A lattice quantizer is a map, ${Q_{\Lambda}: \mathbb{R}^n \rightarrow  \Lambda}$, that sends
 a point, $\mathbf{s}$, to the nearest lattice point in Euclidean distance. That is,
 \begin{equation}
     Q_{\Lambda}(\mathbf{s})=\text{argmin}_{\lambda\in\Lambda}\norm{\mathbf{s-\lambda}}.
 \end{equation}
\end{definition}
\begin{definition}[Voronoi Region]
The \textit{fundamental Voronoi region}, $\mathcal{V}$, of a lattice, $\Lambda$, is the set of all points in $\mathbb{R}^n$ that are closest to the zero vector compared to any other lattice point. That is, 
${\mathcal{V}=\{\mathbf{s}: Q_{\Lambda}(\mathbf{s})=0\}}$.
\end{definition}
}
\textcolor{black}{
\begin{definition}[Second Moment of a Lattice]\label{def:SecondMoment}
The \textit{second moment} of the lattice $\Lambda$ is defined as the second moment per dimension of a random variable $\mathbf{U}$ which is uniformly distributed over the Voronoi region $\mathcal{V}$:
\begin{equation}
    \sigma^2(\Lambda) = \frac{1}{n}E\left[\norm{\mathbf{U}^2}\right]= \frac{1}{\text{Vol}(\mathcal{V})}\cdot \frac{1}{n}\int_{\mathcal{V}} \| \mathbf{x} \|^2 \, d\mathbf{x}.
\end{equation}
\end{definition}
\begin{definition}[Modulus] Let $[\mathbf{s}] \text{mod} \ \Lambda$ denote the quantization error of $\mathbf{s}\in\mathbb{R}^n$ with respect to the lattice $\Lambda$. That is,
\begin{equation}
[\mathbf{s}] \text{mod} \ \Lambda =\mathbf{s}-Q_{\Lambda}(\mathbf{s}).
\end{equation}  
\end{definition}
For all $\mathbf{s,t} \in \mathbb{R}^n$ and $\Lambda_c \subseteq \Lambda_f$, the $\text{mod} \ \Lambda$ operation satisfies:
\begin{align}
\label{eq:dist}
&[\mathbf{s} + \mathbf{t}] \text{mod} \ \Lambda =
\big[[\mathbf{s}] \text{mod} \ \Lambda +\mathbf{t}\big] \text{mod} \ \Lambda 
\\&
\label{eq:QuantizationMod}
\hspace{-0.1cm}\left[Q_{\Lambda_f}(\mathbf{s})\right]\ \text{mod} \ \Lambda_c =
\left[Q_{\Lambda_f}([\mathbf{s}]\ \text{mod} \ \Lambda_c)\right]\ \text{mod} \ \Lambda_c 
\\&
\label{eq:Zscal}
[a\mathbf{s}] \text{mod} \ \Lambda =
[a[\mathbf{s}] \text{mod} \ \Lambda]\text{mod} \ \Lambda \quad \forall a\in \mathbb{Z} 
\\&
\label{eq:scal}
\beta [\mathbf{s}] \text{mod} \ \Lambda =
[\beta \mathbf{s}] \text{mod} \ \beta \Lambda \quad \forall \beta\in \mathbb{R} 
\end{align}
}
\textcolor{black}{
A nested lattice code is a lattice code whose bounding region is the Voronoi region of a sub-lattice. 
Formally, let $\Lambda_c$ and $\Lambda_f$, be a pair of $n$-dimensional lattices with Voronoi regions $\cV_c$ and $\cV_f$, respectively, such that $\Lambda_c$ is a subset of $\Lambda_f$, i.e., $\Lambda_c \subset \Lambda_f$.
Usually, $\Lambda_c$ and $\Lambda_f$ are called the coarse and the fine lattice, respectively.
The nested lattice code is thus given by, $\cC=\{\Lambda_f \cap \cV_c\}$, and its rate is equal to \cite{zamir2014lattice},
\begin{equation}\label{equ-lattice rate}
R=\frac{1}{n}\log|\cC|=\frac{1}{n}\log|\Lambda_f \cap \cV_c|=\frac{1}{n}\log|p^L|=\frac{L\log{p}}{n}.
\end{equation}
The lattices should be chosen appropriately with respect to $p$, which grows together with the dimension $n$ to ensure desired properties such as AWGN good \cite{zamir2014lattice}.
}

%% file: figs/PIR_AWGN_MAC_Model.tex
\begin{figure}[!t] 
\center
\begin{tikzpicture}[scale=0.6]
\tikzmath{\x1 = 0; \y1 =0; 
\x2 = \x1 + 3; \y2 =\y1 +5; } 
\shade[ rounded corners,
        left color= yellow!80,
        right color= red!50,
        shading angle = 45
        ] (\x1,\y1) rectangle (\x2,\y2); 
\node at (\x1/2 + \x2/2,\y2 + 0.5) {\texttt{1}};
\draw[rounded corners, black, thick] (\x1+0.1,\y2-1) rectangle (\x2-0.1,\y2-0.1)  node[pos=.5] {$W_1$};
\draw[rounded corners, black, thick] (\x1+0.1,\y2-2) rectangle (\x2-0.1,\y2-0.1-1)  node[pos=.5] {$W_2$}; 
\foreach \z in {0,-0.5,-1}{
\filldraw [black] (\x1/2+\x2/2,\y2-2.5 + \z) circle (1.5pt);
}
\draw[rounded corners, black, thick] (\x1+0.1,\y1+0.1) rectangle (\x2-0.1,\y1-0.1+1)  node[pos=.5] {$W_M$};
\tikzmath{\x1 = \x2 + .5; \y1 =0; 
\x2 = \x1 + 3; \y2 =\y1 +5; } 
\shade[ rounded corners,
        left color= yellow!80,
        right color= red!50,
        shading angle = 45
        ] (\x1,\y1) rectangle (\x2,\y2);
\node at (\x1/2 + \x2/2,\y2 + 0.5) {\texttt{2}};
\draw[rounded corners, black, thick] (\x1+0.1,\y2-1) rectangle (\x2-0.1,\y2-0.1)  node[pos=.5] {$W_1$};
\draw[rounded corners, black, thick] (\x1+0.1,\y2-2) rectangle (\x2-0.1,\y2-0.1-1)  node[pos=.5] {$W_2$}; 
\foreach \z in {0,-0.5,-1}{
\filldraw [black] (\x1/2+\x2/2,\y2-2.5 + \z) circle (1.5pt);
}
\draw[rounded corners, black, thick] (\x1+0.1,\y1+0.1) rectangle (\x2-0.1,\y1-0.1+1)  node[pos=.5] {$W_M$};
\foreach \z in {0,0.5,1}{
\filldraw [black] (\x2+0.5+\z,\y2-3) circle (1.5pt);
}
\tikzmath{\x1 = \x2 + 2; \y1 =0; 
\x2 = \x1 + 3; \y2 =\y1 +5; } 
\shade[ rounded corners,
        left color= yellow!80,
        right color= red!50,
        shading angle = 45
        ] (\x1,\y1) rectangle (\x2,\y2);
\node at (\x1/2 + \x2/2,\y2 + 0.5) {\texttt{N}};
\draw[rounded corners, black, thick] (\x1+0.1,\y2-1) rectangle (\x2-0.1,\y2-0.1)  node[pos=.5] {$W_1$};
\draw[rounded corners, black, thick] (\x1+0.1,\y2-2) rectangle (\x2-0.1,\y2-0.1-1)  node[pos=.5] {$W_2$}; 
\foreach \z in {0,-0.5,-1}{
\filldraw [black] (\x1/2+\x2/2,\y2-2.5 + \z) circle (1.5pt);
}
\draw[rounded corners, black, thick] (\x1+0.1,\y1+0.1) rectangle (\x2-0.1,\y1-0.1+1)  node[pos=.5] {$W_M$}; 
\shade[rounded corners,
        left color= lightgray,
        right color= blue!30,
        shading angle = 0
        ] (2.75,\y1-3) rectangle (8.75,\y1-2)
        node[pos=0.5] {\texttt{User}};
\draw[rounded corners, blue, very thick] (2.75,\y1-3) rectangle (8.75,\y1-2);
\tikzmath{\x1 = 0; \y1 =0; 
\x2 = \x1 + 3; \y2 =\y1 +5; } 
\draw[black, thick] (\x2/2,\y1) -- node[midway,right] {$h_1$}(\x2/2,\y1-1) -- (\x2/2+4.25,\y1-1);
\draw[black, thick] (\x2/2+3.5,\y1) -- node[midway,right] {$h_2$}(\x2/2+3.5,\y1-1);
\draw[black, thick] (\x2/2+8.5,\y1) -- node[midway,right] {$h_N$}(\x2/2+8.5,\y1-1) -- (\x2/2+8.5-4.25,\y1-1);
\draw[black, thick, ->] (5.75,\y1-1) -- (5.75,\y1-1.3);
\draw[black, thick, ->] (5,\y1-1.5) node[align=left,]{\hspace{-1.5em}$Z$} -- (5.55,\y1-1.5) ;
\draw[black, thick] (5.75,\y1-1.5) circle (0.2cm);
\draw[black, thick] (5.62,\y1-1.5) -- (5.88,\y1-1.5);
\draw[black, thick] (5.75,\y1-1.5 + 0.13) -- (5.75,\y1-1.5 - 0.13);
\draw[black, thick, ->] (5.75,\y1-1.7) -- (5.75,\y1-2);

\end{tikzpicture}
\caption{System model of $N$ databases connected to a user via a block-fading AWGN MAC.}
\label{fig:PIR_AWGN_MODEL}
\end{figure}
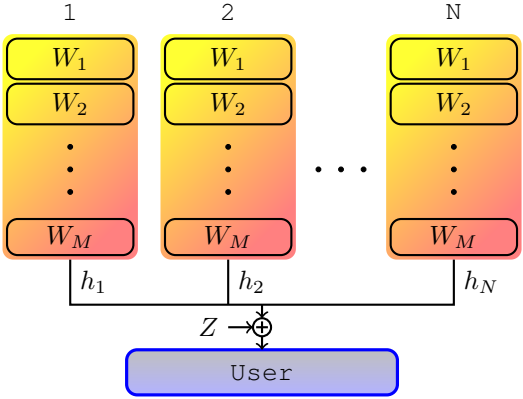

%% file: Joint_FadingNew.tex
\section{Main Results}
The following Theorem presents an achievable rate for the PIR problem over
block-fading AWGN-MAC,

\begin{theorem}\label{theorem:JointFadingNew}
Consider the PIR problem with $N\geq2$ databases over a block-fading AWGN. 
Then, for any non-empty subsets of databases $\mathcal{S}_1, \mathcal{S}_2$ satisfying ${\mathcal{S}_1\cap \mathcal{S}_2=\emptyset}$,
$\mathcal{S}_1 \cup \mathcal{S}_2 \subseteq\{1, ..., N\}$, the following PIR rate is achievable,
\begin{equation}\label{eq:fading_rate_new}
R^{eq}_{PIR}=\frac{1}{2}\log^{+}\left(\frac{1}{2}+\Tilde{h_1}^2P\right),
\end{equation}
where $\Tilde{h}_i\overset{\Delta}{=}\sum_{k\in \mathcal{S}_i} \left|{h_k}\right|$, $ \Tilde{h}_1 \leq \Tilde{h}_2$.
\end{theorem}

Interestingly, unlike the classical PIR problem, the achievable rate is independent of the number of messages $M$.
In order to analyze the achievable PIR rate in Theorem \ref{theorem:JointFadingNew}, one should note that the user may choose $\mathcal{S}_1$ and $\mathcal{S}_2$ to maximize $R_{PIR}^{eq}$. 
Namely, we have the following optimization problem,
$$
\max_{\substack{\mathcal{S}_1,\mathcal{S}_2 \\
\Tilde{h}_1\leq\Tilde{h}_2}}
\left\{\frac{1}{2}\log^{+}\left(\frac{1}{2}+\left(\sum_{k\in \mathcal{S}_1} h_k\right)^2P\right)\right\}.
$$
Thus, one has to choose $\mathcal{S}_1$ which maximize $\Tilde{h}_1$ such that $\Tilde{h}_1\leq\Tilde{h}_2$, i.e., find $\mathcal{S}_1$ and $\mathcal{S}_2$ which bring $\Tilde{h}_1$ as close as possible to  $\Tilde{h}_2$.
Finding the optimal solution is related to the subset sum problem, known to be NP-hard. 
Hence, in Theorem \ref{theorem:LowerBoundExpected} below, we propose a low-complexity suboptimal solution that provides a lower bound for the achievable rate.
Yet, to put Theorem \ref{theorem:JointFadingNew} in context, we first consider the gap from the AWGN MAC sum-rate capacity, which does not assume any privacy constraint.
The following lemma, whose proof is given in Appendix \ref{Appendix:Gap}, shows that the achievable PIR rate is asymptotically optimal with respect to the unrestricted channel capacity for $N$ large enough.

\begin{lemma}\label{lemma:Gap}
    The PIR rate with $N$ databases over a block-fading AWGN MAC given in Theorem \ref{theorem:JointFadingNew} is asymptotically optimal with respect to the unrestricted channel capacity for $N$ large enough. That is,
    \begin{equation}
        C_{SR}^{\textcolor{black}{MAC}}-\max_{\substack{\mathcal{S}_1,\mathcal{S}_2 \\
\Tilde{h}_1\leq\Tilde{h}_2}}\left\{R_{PIR}^{eq}\right\} \leq 1+O\left(\frac{1}{N}\right).
    \end{equation}
\end{lemma}

Thus, Lemma \ref{lemma:Gap} asserts that the price paid for enforcing the privacy constraint is limited by only $1$ bit when the number of databases is large enough.
\textcolor{black}{
This should be compared to the $C\&F$ based scheme suggested in \cite{shmuel2021private}, which achieves a similar result only in the no-fading scenario.
With block fading and using the $C\&F$ based scheme of \cite{shmuel2021private} there is a mismatch between the partitioned fading coefficients $\Tilde{h}_1, \Tilde{h}_2$ and the desired integer coefficients one tries to decode. 
This mismatch results in a substantial rate loss.
More details about it will be given in Section \ref{sec:compare}.}

\subsection{Symmetric PIR}
\textcolor{black}{For the classical SPIR setting (noiseless and orthogonal channels \cite{sun2018capacity}), it has been shown that having common randomness $S$ is essential \cite{gertner1998protecting}.}
However, requiring common randomness among the databases is not a trivial matter.
Hence, characterizing the size of $S$ is of interest.
We define the  amount of common randomness relative to the message size and denote it by $\rho$ \cite{sun2018capacity}, i.e.,
$$\rho = \frac{H(S)}{H(W_i)}=\frac{H(S)}{L\log(p)}.$$

Next, we present the achievable SPIR rate over the block-fading AWGN MAC.
\begin{theorem}\label{theorem:SPIR}
Consider the SPIR problem with $N\geq2$ databases over a block-fading AWGN.
Then, for $\rho\geq1$, and for any non-empty subsets of databases $\mathcal{S}_1, \mathcal{S}_2$ satisfying ${\mathcal{S}_1\cap \mathcal{S}_2=\emptyset}$,
$\mathcal{S}_1 \cup \mathcal{S}_2 \subseteq\{1, ..., N\}$, the following PIR rate is achievable,
\begin{equation}\label{eq:fading_rate_SPIR}
R^{eq}_{SPIR}=\frac{1}{2}\log^{+}\left(\frac{1}{2}+\Tilde{h_1}^2P\right) ,
\end{equation}
where, $\Tilde{h}_i\overset{\Delta}{=}\sum_{k\in \mathcal{S}_i} \left|{h_k}\right|$, $ \Tilde{h}_1 \leq \Tilde{h}_2$.
\end{theorem}

The following observations place Theorem \ref{theorem:SPIR} in perspective.
\begin{enumerate}
    \item It is well established that in the classical setting, $C_{PIR}>C_{SPIR}$ \cite{sun2018capacity}.
    Nevertheless, according to Theorem \ref{theorem:SPIR}, adding the symmetric constraint requires some common randomness at the databases, yet this does not introduce any rate penalty compared to Theorem \ref{theorem:JointFadingNew}, that is     ${R^{eq}_{PIR}=R^{eq}_{SPIR}}$.
    \item Duo to Lemma \ref{lemma:Gap} and the fact that $R^{eq}_{PIR}=R^{eq}_{SPIR}$, we conclude that $R^{eq}_{SPIR}$ is also asymptotically optimal with respect to the unconstrained AWGN MAC sum-rate capacity. 
\end{enumerate}

\textcolor{black}{
In a practical scenario, requiring databases to have common randomness could be challenging.
In the following theorem, we provide the result of a novel joint SPIR-channel coding scheme over an AWGN MAC without fading, which achieves symmetric privacy without assuming any shared randomness among the servers.}
This result is not trivial, as we know it is not feasible in the classic setting (orthogonal channels and non-communicating servers \cite{gertner1998protecting}).

\begin{theorem}\label{theorem:SPIRNoKey}
Consider the SPIR problem over an AWGN MAC with $N=2$ replicated databases, each containing $M$ messages.
Then, the following SPIR rate is achievable,
\begin{equation}
R_{SPIR}=\frac{1}{2} \log\left(\frac{4P}{M}\right).
\end{equation}
\end{theorem}
\textcolor{black}{
It is essential to note that the achievable rate in Theorem~\ref{theorem:SPIRNoKey} decreases with the number of messages $M$, which can significantly limit performance in settings involving large databases. Nonetheless, the scheme offers a distinct advantage in terms of simplicity and practical implementation, as it does not rely on any shared randomness among the databases.
This requirement can be difficult to realize in distributed systems.
}


\begin{figure*}[t] 
    \centering
    \begin{subfigure}[t]{0.49\linewidth}
             \centering            
             \includegraphics[width=0.7\linewidth]{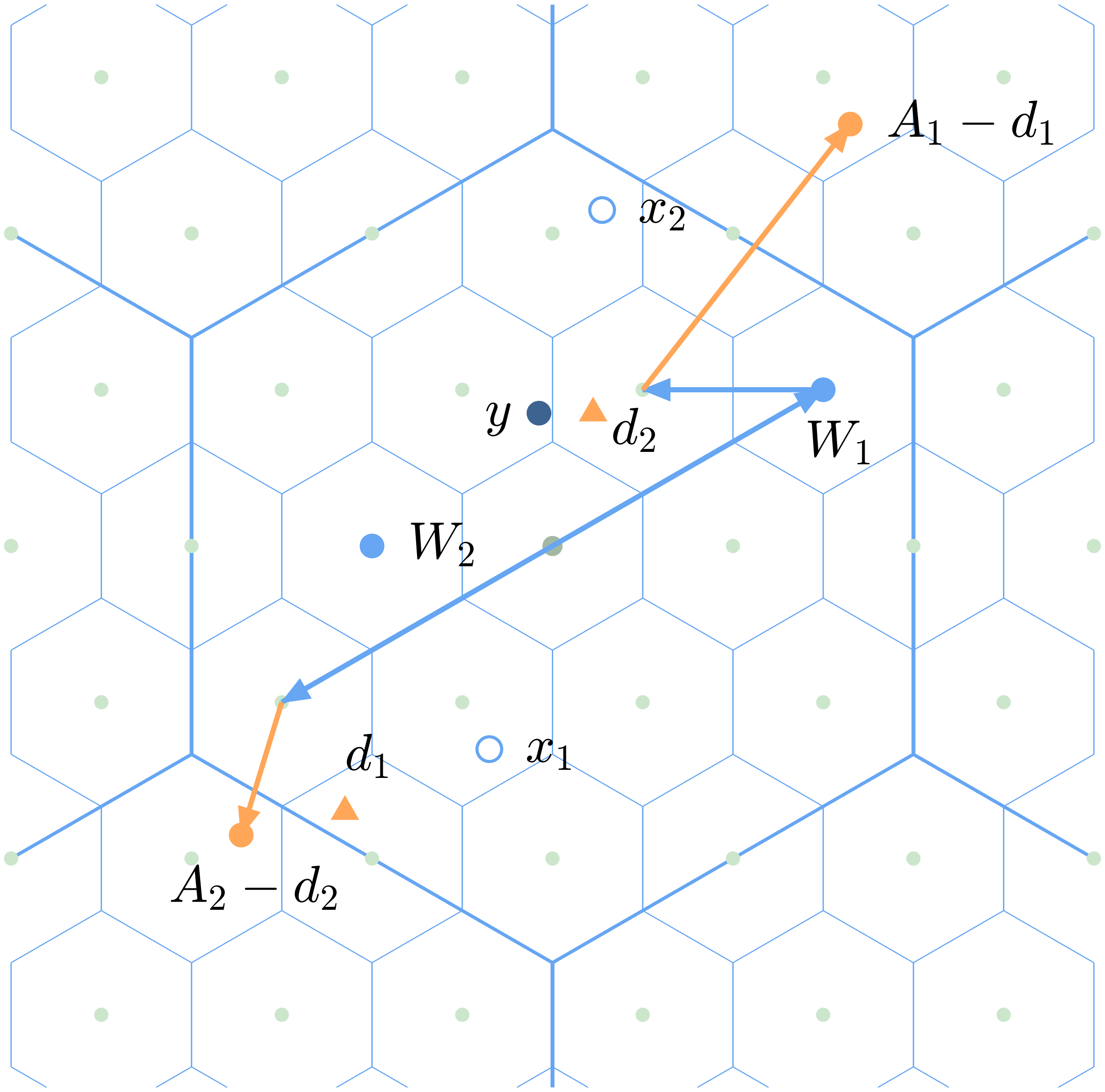}
             \caption{}
    \end{subfigure}
    \begin{subfigure}[t]{0.49\linewidth}
             \centering
             \includegraphics[width=0.7\linewidth]{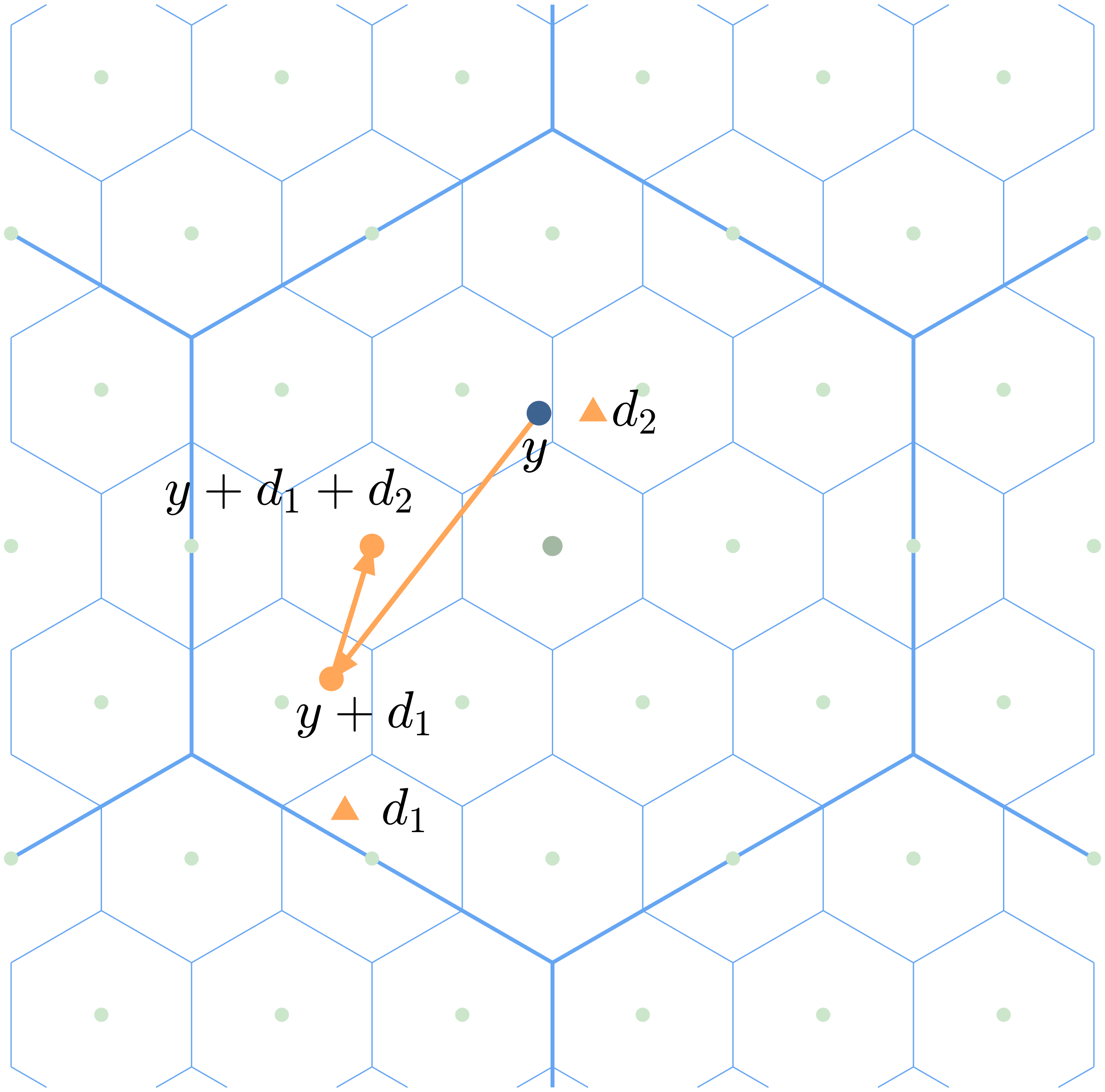}
             \caption{}
    \end{subfigure}
    \caption{\textcolor{black}{An illustration of a simplified scheme described in the proof of Theorem\ref{theorem:JointFadingNew}, assuming $N=2$ databases, each storing the same $M=2$ messages, and a channel without fading. 
    The codewords lie in the fine lattice confined within the coarse lattice’s Voronoi region. 
    $W_1$ and $W_2$ represent segments of the full messages stored in the databases (\textcolor{black}{blue filled} dots), while the dithers are represented by orange triangles.
    (a) Encoding stage: To privately retrieve $W_2$, the user generates the queries $Q_1(2) = [1 \ 1]$ and $Q_2(2) = [-1 \ 0]$, following \eqref{eq:fading_queries}. 
    This results in $\mathbf{A}_1 = W_1+W_2$ and $\mathbf{A}_2=-W_1$.
    The transmitted codewords $\mathbf{x}_1$ and $\mathbf{x}_2$ (\textcolor{black}{blue circles}) are obtained by adding dithers and applying a modulo operation to enforce the power constraint. 
   The user receives the superimposed signal $\mathbf{y}$.
    (b) Decoding stage: The user computes $\hat{W}_2 = [\mathbf{y} + \b{d}_1 + \b{d}_2] \mod \Lambda_c$, successfully recovering the desired message $W_2$.}}
     \label{fig:illustration}
\end{figure*}

\section{A PIR Scheme For the Block-Fading AWGN-MAC}\label{sec:Achievability}
\textcolor{black}{ In this section we present the achievability scheme resulting Theorem \ref{theorem:JointFadingNew}.}

\subsection{Coding Scheme}
Our scheme utilizes nested lattice codebooks constructed as in \cite[Section 4.B]{nazer2011compute}, using two $n$-dimensional lattices $\Lambda_c \subset \Lambda_f$ with Voronoi regions $\cV_c$ and $\cV_f$, respectively.
The nested lattice codebook is then given by $\cC=\{\Lambda_f \cap \cV_c\}$ and is known to both the user and the databases.
By \cite[Lemmas 4-6]{nazer2011compute}, there exists an isomorphism $\phi(\cdot)$ between $\mathbb{F}^l_p$ and the codebook $\cC$  where $L=kl$, and $k,l\in\mathbb{N}$, namely:
$$\b{s} = (s_1,...,s_l)\in\mathbb{F}^l_p \mapsto \boldsymbol{\lambda}=(\lambda_1,...,\lambda_n)\in\cC.$$

In essence, within our proposed scheme, each database encodes its answer using the same nested lattice code with a rate $R$ that will be determined. 
This encoding ensures that the answers can be added constructively to retrieve the requested message, while each answer, separately, remains independent of the desired index message.

The queries and the assignment to whom they are being sent differ and depend on the channel vector $\mathbf{h}$, determined by nature.
Thus, Theorem \ref{theorem:JointFadingNew} initially presents the result for a fixed $\mathbf{h}$ and fixed subsets of users. 


\begin{IEEEproof}[Proof of Theorem \ref{theorem:JointFadingNew}]\label{the:th1 proof}
Assume the user wants to retrieve message $W_i$ privately.\\
\textcolor{black}{\textit{Query:}} 
To do that, the user generates a random vector $\mathbf{b}$ of length $M$ such that each entry is either $1$ or $0$, independently and with equal probability.
Then, the user divides the databases into two non-intersecting subsets, denoted as $\mathcal{S}_1$ and $\mathcal{S}_2$, for which he sends the query $Q_1(i)$ to each member in $\mathcal{S}_1$ and $Q_2(i)$ to each member in $\mathcal{S}_2$. 
The queries are given as follows
\begin{equation} \label{eq:fading_queries}
    \begin{split}
        & Q_1(i)=\mathbf{b}, \quad Q_2(i)=-\mathbf{b}-\mathbf{e}_i \quad \text{if } b_i = 0 \\
        & Q_1(i)=\mathbf{b}, \quad Q_2(i)=-\mathbf{b}+\mathbf{e}_i \quad \text{if } b_i = 1
    \end{split}
\end{equation}

\textcolor{black}{Thus, $Q_1\in\{0,1\}^M$ and $Q_2\in\{-1,0\}^M$.
From the databases' perspective, each sees a uniform random vector.}  
\\
\textcolor{black}{\textit{Answers:}} 
\textcolor{black}{
Upon receiving the queries, the databases construct their responses by computing linear combinations of the messages, where the combining coefficients are determined by the query entries. 
That is,
\begin{align}\label{eq:fading_answers}
&\b{A}_k=\sum_{m=1}^M Q_{k,m}(i)W_m,
\end{align}
where $Q_{k,m}(i)$ is the $m$th entry of the vector $Q_{k}(i)$, $k\in\{1,2\}$.
We note that $A_k=(a_k^{(1)},\dots,a_k^{(L)})\in \mathbb{F}_p^L$, and the scheme is focused on the transmission of a single symbol $a_k^{(m)}$, where $1\leq m \leq L$, from each answer.
To construct the entire message, the databases must iterate this process across all $L$ symbols.}

\textcolor{black}{
Without loss of generality, assume $m=1$, i.e., the databases wish to transmit the first symbol from each answer, that is, $a_1^{(1)} $ and $a_2^{(1)}$.
Note that $\b{A}_1+\b{A}_{2}$ is equal to either $W_i$, or $-W_i$. 
This depends on the sign of $b_i$, which is known to the user.
In the same way $a_1^{(m)}+a_2^{(m)} = \pm W_i^{(m)}$.
To encode the symbols, each database maps the relevant entry of the answer to the codebook as follows: $\boldsymbol{\lambda}_1 = \boldsymbol{\phi}(a_1^{(1)})$ and  $\boldsymbol{\lambda}_2 = \boldsymbol{\phi}(a_2^{(1)})$.}

Define $\Tilde{h}_i\overset{\Delta}{=}\sum_{k\in \mathcal{S}_i} \left|{h_k}\right|$.
Assume without loss of generality,  $\Tilde{h}_1\leq\Tilde{h}_2$, and let $\b{d}_1$ and $\b{d}_2$ be two mutually independent dithers which are uniformly distributed over the Voronoi region $\cV_c$. The dithers are known to both the user and the databases. Then, each database transmits
 either $\mathbf{x}_{1}$ or $\mathbf{x}_{2}$ according to the subsets $\mathcal{S}_1$ and $\mathcal{S}_2$ respectively, namely:
\begin{equation} \label{eq:fading_trans}
        \begin{split}
            \mathbf{x}_{1}&=[\boldsymbol{\lambda}_{1}-\mathbf{d}_1]\text{mod} \ \Lambda_c,  \\
            \mathbf{x}_{2}&=\frac{\Tilde{h}_1}{\Tilde{h}_2}\mathbf{x'}_2=\frac{\Tilde{h}_1}{\Tilde{h}_2}[\boldsymbol{\lambda}_{2}-\mathbf{d}_{2}]\text{mod} \ \Lambda_c.
        \end{split}
\end{equation}

\textcolor{black}{Using two dithers allows the distribution of $\bx_{1}$ and $\bx'_{2}$ to be uniform over the Voronoi region, ensuring their independence with $\boldsymbol{\lambda}_{1}$ and $\boldsymbol{\lambda}_{2}$ as well as from each other \cite[Lemma 1]{erez2004achieving}.}
We note that the user may add additional information to the query, informing the database which group the database belongs to and the factor to be multiplied before transmission. 
As a result, the database can control the sign of each coefficient to ensure proper summation.
This additional information does not affect the privacy constraint as shown in \cite{shmuel2021private}.

The received signal by the user  is then given by:
$$\mathbf{y}=\sum_{k\in\mathcal{S}_1} |h_k|\mathbf{x}_{1} +\sum_{k\in\mathcal{S}_21} |h_k|\mathbf{x}_{2} + \mathbf{z} = \Tilde{h}_1( \mathbf{x}_{1}+\mathbf{x'}_{2})+\mathbf{z}.$$

\textcolor{black}{\textit{Decoding:}} To decode $\mathbf{v}\overset{\Delta}{=}\boldsymbol{\phi}(W_i^{(1)})$, the user computes the following, 
\begin{equation*}
\hat{\b{v}}=\left[\alpha\frac{1}{\Tilde{h}_1}\mathbf{y}+\mathbf{d}_{1}+\mathbf{d}_{2}\right]\text{mod} \ \Lambda_c,
\end{equation*}  
where $0\leq\alpha\leq1$ will be optimized later.

To compute the expression, we transform the channel to the Modulo-Lattice Additive Noise (MLAN) channel \cite{erez2004achieving} as follows,
\begin{fleqn}[\parindent]
\begin{equation}\label{eq:theorem_Fading_Comput}
\begin{split}
\hat{\b{v}}&=\left[\alpha\frac{1}{\Tilde{h}_1}\mathbf{y}+\mathbf{d}_{1}+\mathbf{d}_{2}\right]\text{mod} \ \Lambda_c 
\\ &\overset{}{=}  
\left[\alpha\frac{1}{\Tilde{h}_1}(\Tilde{h}_1\mathbf{x}_{1}+\Tilde{h}_2\mathbf{x}_{2}+\mathbf{z})+\mathbf{d}_{1}+\mathbf{d}_{2}\right]\text{mod} \ \Lambda_c  
\\ &\overset{}{=}  
\left[\alpha (\mathbf{x}_{1}+\frac{\Tilde{h}_2}{\Tilde{h}_1}\mathbf{x}_{2}+\frac{1}{\Tilde{h}_1}\mathbf{z})+\mathbf{d}_{1}+\mathbf{d}_{2}\right]\text{mod} \ \Lambda_c  
\\ &\overset{}{=}  
\left[\alpha (\mathbf{x}_{1}+\mathbf{x'}_{2}+\frac{1}{\Tilde{h}_1}\mathbf{z})+\mathbf{d}_{1}+\mathbf{d}_{2}\right]\text{mod} \ \Lambda_c 
\\ &\overset{(a)}{=} 
\bigg[\mathbf{x}_{1}+\mathbf{x'}_{2}+(\alpha-1)(\mathbf{x}_{1}+\mathbf{x'}_{2})\\
&\qquad \qquad \qquad \qquad +\alpha \frac{1}{\Tilde{h}_1} \mathbf{z}+ \mathbf{d}_{1}+\mathbf{d}_{2}\bigg]\text{mod} \ \Lambda_c  
\\ &
\overset{}{=} 
\bigg[\left[\boldsymbol{\lambda}_{1}-\mathbf{d}_{1}\right]\text{mod} \ \Lambda_c +\left[\boldsymbol{\lambda}_{2}-\mathbf{d}_{2}\right]\text{mod} \ \Lambda_c\\
& \qquad -(1-\alpha)(\mathbf{x}_{1}+\mathbf{x'}_{2})+\alpha \frac{1}{\Tilde{h}_1} \mathbf{z}+ \mathbf{d}_{1}+\mathbf{d}_{2}\bigg]\text{mod} \ \Lambda_c  
\\ &\overset{(b)}{=} 
\left[\mathbf{v}-(1-\alpha)(\mathbf{x}_{1}+\mathbf{x'}_{2})+\alpha\frac{1}{\Tilde{h}_1} \mathbf{z}\right]\text{mod} \ \Lambda_c 
\end{split}
\end{equation}
\end{fleqn}
where (a) is the  MLAN  equivalent channel.
(b) follows from the distributive property of the $\text{mod} \ \Lambda_c$ operation and due to the structure of the answers \eqref{eq:fading_answers} where we assume that $b_i=1$. In case $b_i=0$, we would result in a negative sign to $\mathbf{v}$. 
\textcolor{black}{However, since $b_i$ is known to the user, the recovered value $\hat{\mathbf{v}}$ can be corrected by multiplying it by $-1$ if necessary.}
Finally, we define the equivalent noise term
${\mathbf{z}_{eq}\triangleq-(1-\alpha)(\mathbf{x}_{1}+\mathbf{x}_{2}')+\alpha\frac{1}{\Tilde{h}_1} \mathbf{z}} $.
\textcolor{black}{As shown in \cite{nazer2011compute}, for sufficiently large $n$, the second moment of $\mathbf{z}_{eq}$ approaches
$ {\sigma^{2}_{eq}=\frac{1}{n}E\left[\norm{\mathbf{z}_{eq}}^2\right] = 2P(1-\alpha)^2+\frac{1}{\Tilde{h}_1^2}\alpha^2 \ ;}$
This follows from the Crypto Lemma and the use of independent dithers.
To minimize $\sigma^2_{eq}$, we optimize over $\alpha$, yielding}
${\alpha_{opt} = \frac{2P}{2P+\frac{1}{\Tilde{h}_1^2}}}$;
${\sigma^{2}_{opt}=\frac{2P\frac{1}{\Tilde{h}_1^2}}{2P+\frac{1}{\Tilde{h}_1^2}}}$. 
\textcolor{black}{With this choice, the decoding error probability can be made arbitrarily small as $n \to \infty$, resulting in the achievable rate $\frac{1}{2}\log\left(\frac{P}{\sigma^2_{opt}}\right)$  \cite{erez2004achieving}.
Substituting yields the final equivalent PIR rate:}
$R^{eq}_{PIR} = \frac{1}{2}\log^{+}\left(\frac{1}{2}+\Tilde{h_1}^2P\right)$.

\begin{figure}[!t] 
    \centering           
    \includegraphics[width=\linewidth]{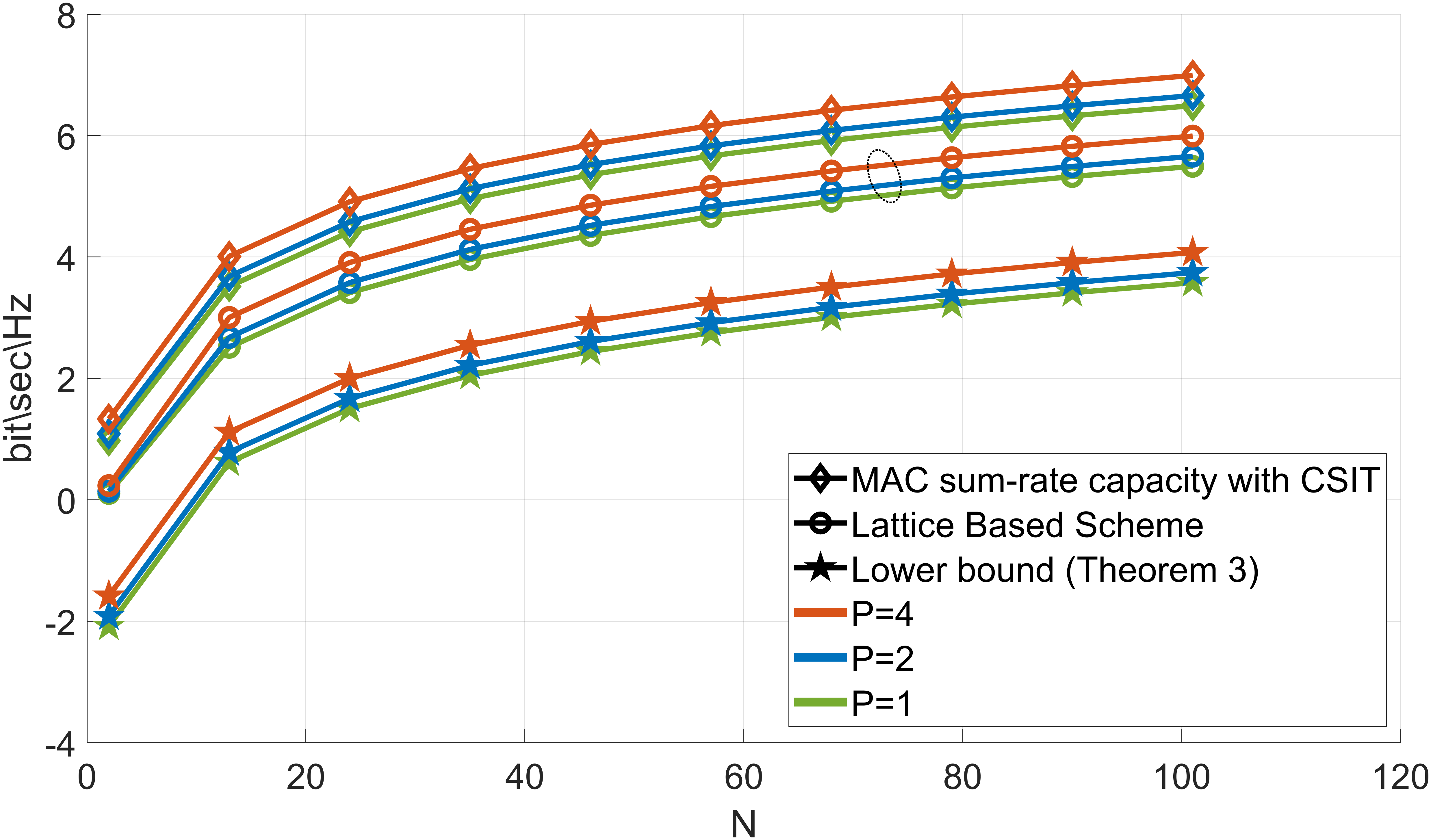}
    \caption{The average PIR rate (solid lines) as a function of $N$.}
    \label{fig:MISOvsLattice_RvsN}
\end{figure}

Next, we show that the user-privacy requirement \eqref{eq:User-privacy} for the $j$th database is fulfilled, while considering the whole $k$ iteration process, namely:
\textcolor{black}{
\begin{equation*}
    \begin{split}
&I(\theta;Q_g(\theta),W_1^M,\mathbf{x}_{g,1}(\theta),\mathbf{x}_{g,2}(\theta),\dots,\mathbf{x}_{g,k}(\theta))
\\&
\overset{(a)}{=}
I(\theta;Q_g(\theta),W_1^M,\mathbf{A}_g(\theta))
\\&
\overset{(b)}{=}I(\theta;Q_g(\theta),W_1^M)=0\\
    \end{split}
\end{equation*}
where $\mathbf{x}_{g,1}(\theta),\mathbf{x}_{g,2}(\theta),\dots,\mathbf{x}_{g,k}(\theta)$ denotes the answers from $k$ sequential iterations from the same database.
(a) is since there is a one-to-one mapping between $\mathbf{x}_{g,1}(\theta),\mathbf{x}_{g,2}(\theta),\dots,\mathbf{x}_{g,k}(\theta)$ to the elements of $\mathbf{A}_g(\theta)$.
(b) holds because $A_g(\theta)$ is a function of $Q_g(\theta),W_1^M$.
The final step follows from the mutual independence of $\theta$, $Q_g(\theta)$, and $W_1^M$.
Note that for any $g\in\{1,2\}$, the query $Q_g(\theta)$ is an $i.i.d.$ $(\frac{1}{2},\frac{1}{2})$ random vector.}
\end{IEEEproof}


\textcolor{black}{Figure \ref{fig:illustration} provides a simplified geometric illustration of the proposed scheme under a no-fading assumption.}

\textcolor{black}{The achievable rate in \eqref{eq:fading_rate_new} scales as $O(\log P)$, indicating its asymptotic optimality relative to the sum-rate capacity of the AWGN MAC when $P$ is sufficiently large. Numerical evaluations, depicted in Figure \ref{fig:MISOvsLattice_RvsN}, based on Monte Carlo simulations, confirm that the achievable PIR rate (circles) maintains a constant gap from the unconstrained AWGN MAC sum-rate capacity with CSIT (diamonds) as $N$ increases.
These findings confirm that the achievable PIR rate is asymptotically optimal with respect to the AWGN MAC sum-rate capacity for both $P$ and $N$, reinforcing its efficiency and potential for practical applications.
Furthermore, we see that the $1$ bit/sec/Hz capacity gap is attained with only a few databases, indicating that near-optimal performance is achievable in practical settings.
}

\begin{figure*}[t] 
    \centering
     \includegraphics[width=0.9\linewidth]{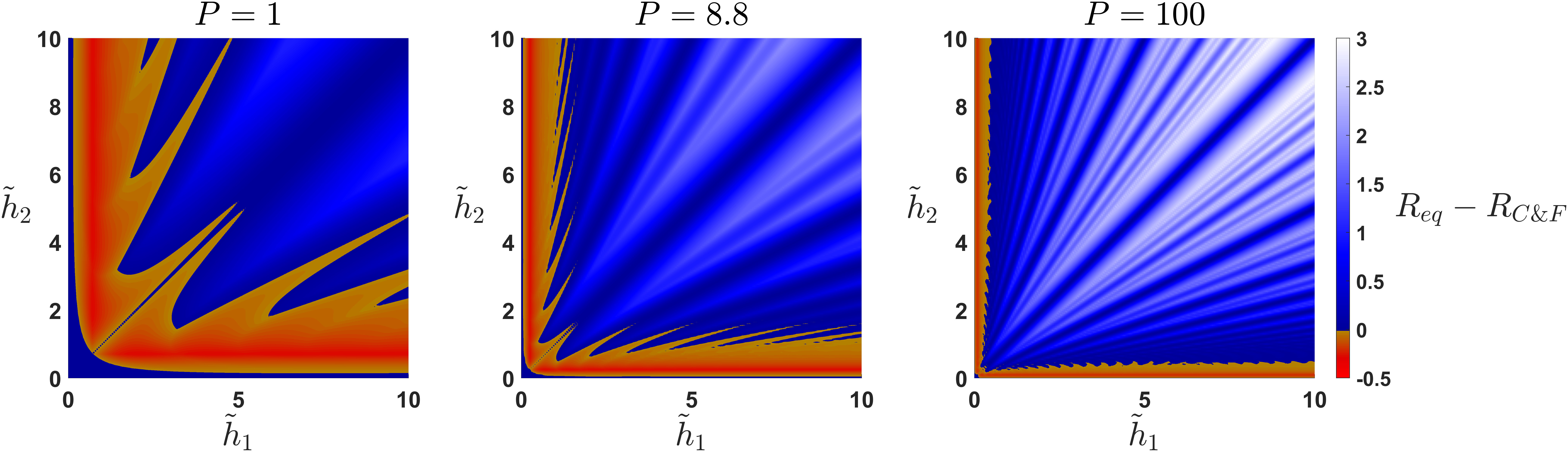}
    \caption{$R^{eq}_{PIR} - R_{PIR}^{C\&F}$, for different $\tilde{h}_1$ and $\tilde{h}_2$.
    Blue and white colors depict positive values, while reds represent negative values.
    Each graph is evaluated for different values of $P$.}
    \label{fig:compare2}
\end{figure*}

\textcolor{black}{In the scheme outlined above, the databases are divided into two groups to optimize the PIR rate. This raises the question of whether dividing into more groups could enhance performance. While we currently lack a converse proof to rule out any such improvements, the analysis of the direct suggests that the two-group structure is nearly optimal.
Nonetheless, exploring different grouping strategies for potential rate improvements remains an interesting topic for future research.}

\subsection{Lower Bound on the Expected Achievable Rate}
To maximize the PIR rate as given by \eqref{eq:fading_rate_new}, it is crucial for the user to carefully choose $\mathcal{S}_1$ and $\mathcal{S}_2$. 
This selection process leads us to a global optimization problem, which can be expressed as:
\begin{equation}\label{eq:optimization}
\max_{\substack{\mathcal{S}_1,\mathcal{S}_2 \\
\Tilde{h}_1\leq\Tilde{h}_2}}
\left\{\frac{1}{2}\log^{+}\left(\frac{1}{2}+\left(\sum_{k\in \mathcal{S}_1} h_k\right)^2P\right)\right\}\textcolor{black}{.}
\end{equation}

\textcolor{black}{
This optimization seeks to maximize the achievable PIR rate by partitioning the databases into two groups such that the sum of channel gains in $\mathcal{S}_1$  is as close as possible to the sum in $\mathcal{S}_2$.
The challenge arises from the channel gains being random variables, making the partitioning a non-trivial combinatorial problem.}

\textcolor{black}{
Next, we derive an asymptotic lower bound on the expected PIR rate in Theorem \ref{theorem:LowerBoundExpected} with respect to channel gains $\mathbf{h}$, to provide analytical insight into the achievable performance.
The proof (found in Appendix \ref{Appendix:LowerBound}) employs a low-complexity, sub-optimal method to partition and demonstrates that even a simple selection strategy achieves a rate close to the achievable PIR rate.}

\begin{theorem}\label{theorem:LowerBoundExpected}
 The expected PIR rate in \eqref{eq:optimization} is asymptotically lower-bounded by,
\begin{equation}\label{eq:lowerbound1}
\begin{split}
\mathbb{E}\left[R_{PIR}^{eq,max}\right]&= 
\mathbb{E}\left[\max_{\substack{\mathcal{S}_1,\mathcal{S}_2 \\ \Tilde{h}_1\leq\Tilde{h}_2}} \left\{\frac{1}{2}\log^{+}\left(\frac{1}{2}+\left(\sum_{k\in \mathcal{S}_1} h_k\right)^2P\right)\right\}\right]
\\ &\geq \frac{1}{2}\log\left(\frac{2+N^2Pc}{4}\right)-o(1)
\end{split}
\end{equation}
where $c=\left(\sqrt{\frac{2}{\pi}}-\frac{1}{2}\right)^2$, $o(1)\rightarrow 0$ as $N\rightarrow\infty$.
\end{theorem}

Note that the lower bound on the expected PIR rate given by Theorem \ref{theorem:LowerBoundExpected} scales similarly to the AWGN MAC sum-rate capacity with CSIT \eqref{equ-MISO sum capacity with per-antenna power constraint} with respect to both $P$ and $N$.

%% file: ComparisonFadingSchemes.tex

\section{Comparison Between the Schemes}\label{sec:compare}

We emphasize the key technical distinction between our scheme and the one presented in \cite{shmuel2021private}.
In \cite{shmuel2021private}, the objective is to decode an integer linear combination of the transmitted database responses, aiming for the closest approximation to the actual linear combination received by the user, specifically, 
$\tilde{h_1}\b{x}_1+\tilde{h_2}\b{x}_2 + \b{z}$. 
The decoding stage leverages results from  \cite{nazer2011compute,erez2004achieving}.
In contrast, our proposed scheme adopts a different strategy. 
We design the database responses to ensure their \emph{gains balance}. 
This approach eliminates the restriction to integer linear combinations, thus eliminating the need for the C\&F protocol.

\textcolor{black}{To assess the performance of our scheme, we compare the achievable PIR rate $R^{eq}_{PIR}$ with the C\&F-based rate provided in \cite[Theorem 4]{shmuel2021private}, denoted as}
\begin{equation}\label{theorem:Fading}
R^{C\&F}_{PIR}=\frac{1}{2}\log^{+}\left(\frac{1+P\left(\Tilde{h}_1^2+\Tilde{h}_2^2\right)}{\|\b{a}\|^2+P(a_1\Tilde{h}_2-a_2\Tilde{h}_1)^2}\right),
\end{equation}
where $\b{a}=(a_1,a_2)\in\mathbb{Z}^2$. \textcolor{black}{While both rates exhibit favorable scaling with the number of databases $N$, a key distinction arises in their behavior with respect to the transmit power $P$.
As illustrated in Figure~\ref{fig:compare2}, $R^{C\&F}_{PIR}$ does not scale efficiently with increasing $P$, whereas $R^{eq}_{PIR}$ achieves optimal scaling. 
In particular, as $P$ increases (comparing from the left to the right subplots), $R^{eq}_{PIR}$ outperforms $R^{C\&F}_{PIR}$ across most of the parameter space.
It is also worth noting that the edge regions of the figure correspond to rare channel conditions.
as $N$ increases, the effective gains $\Tilde{h}_1$ and $\Tilde{h}_2$ tend to equalize, making the ratio $\Tilde{h}_1 / \Tilde{h}_2$ approach $1$.}

\begin{figure*}[t] 
    \centering
    \begin{subfigure}[t]{0.49\linewidth}
             \centering            
             \includegraphics[width=\linewidth]{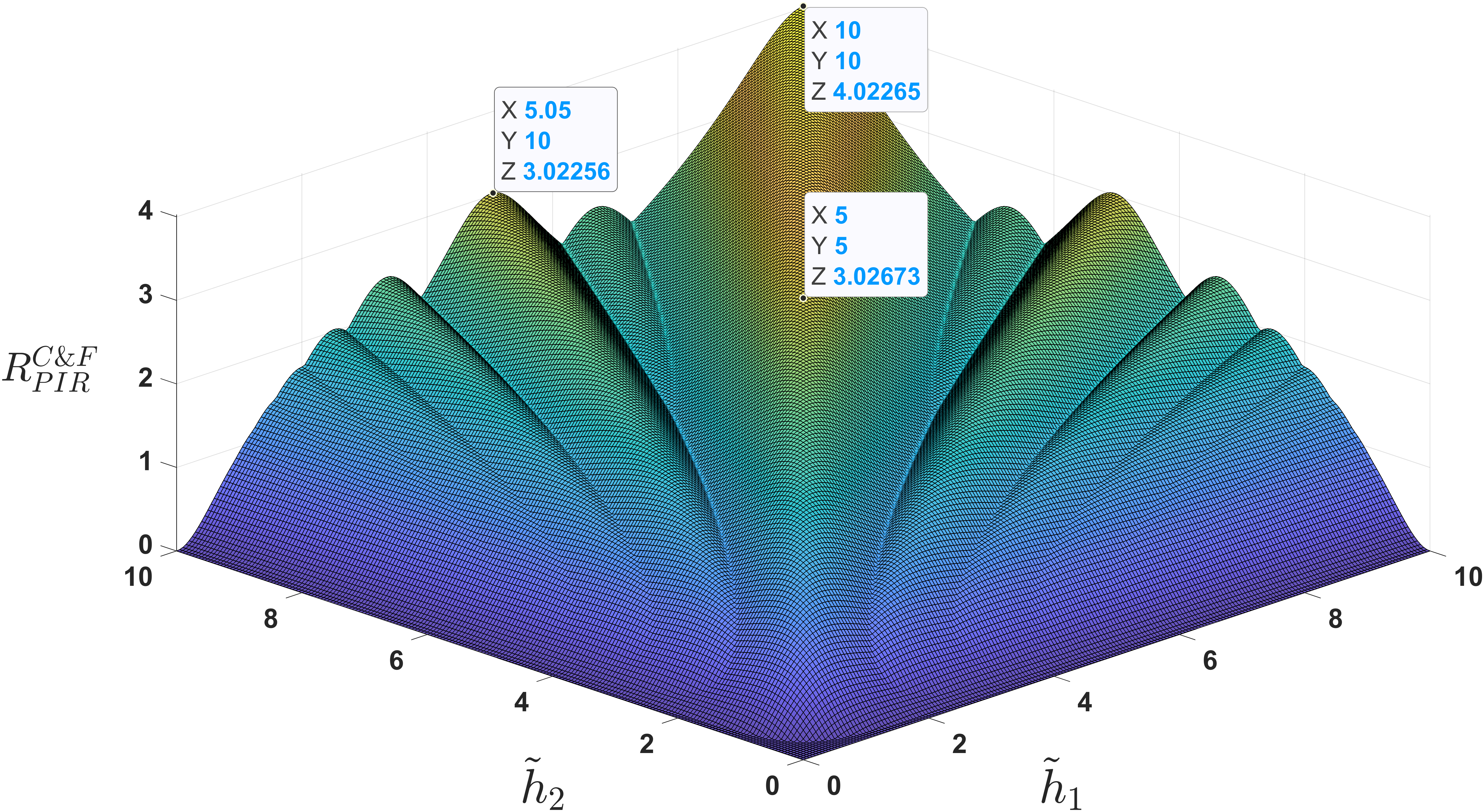}
             \caption{}
             \label{fig:compare1-a}
    \end{subfigure}
    \begin{subfigure}[t]{0.49\linewidth}
             \centering
             \includegraphics[width=\linewidth]{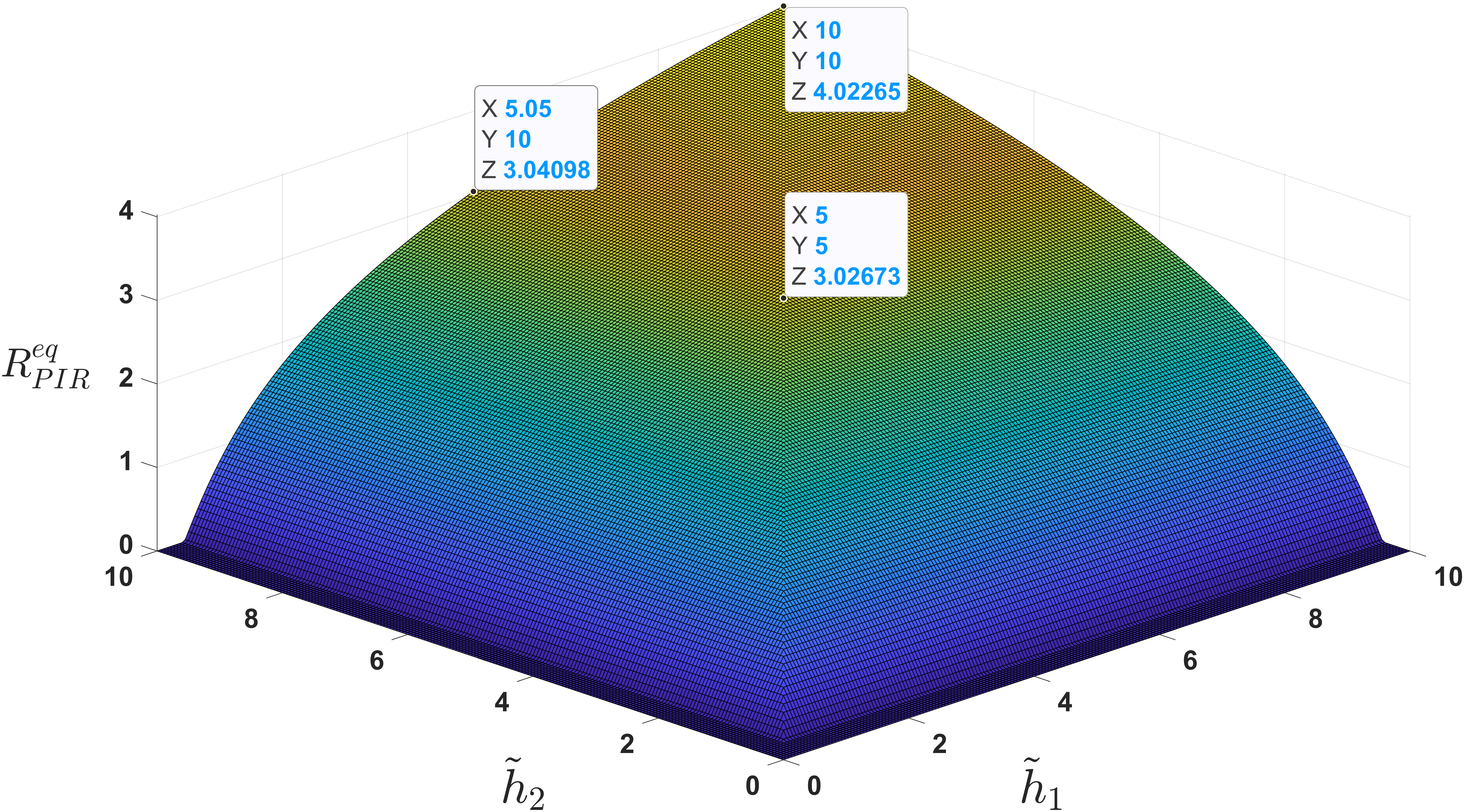}
             \caption{}
             \label{fig:compare1-b}
    \end{subfigure}
    \caption{The graphs shows the rates for different $\Tilde{h_1}$ and $\Tilde{h_2}$. graph (a) is for $R^{C\&F}_{PIR}$ and graph (b) for $R^{eq}_{PIR}$ where P=5.}
     \label{fig:compare1}
\end{figure*}

\textcolor{black}{Further insights are provided in Figure~\ref{fig:compare1}, which shows the achievable rates under varying channel coefficients. This figure reveals a significant degradation in $R^{C\&F}_{PIR}$ across much of the channel space, particularly away from the line $\Tilde{h}_1 = \Tilde{h}_2$. In contrast, $R^{eq}_{PIR}$ exhibits a more robust and monotonic behavior, further highlighting its advantage under general channel conditions.}

\begin{remark}
The inefficiency in the scaling of $R^{C\&F}_{PIR}$ with $P$ arises from a non-zero term in its denominator that is multiplied by $P$.
This term can only be reduced to zero when $\b{a}$ and $\b{\tilde{h}}$ are aligned in the same direction.
Given that the elements of $\b{\tilde{h}}$ are continuous random variables, this scenario is highly unlikely.
\textcolor{black}{To illustrate this, consider a simple example with two databases, where the channel gains are $h_1=0.8$ and $h_2=1.2$.
The received signal is given by
$y=0.8x_1+1.2x_2+z$.
The C\&F scheme attempts to decode an integer linear combination of the transmitted symbols. 
A reasonable choice of integer coefficients is $a=(1,1)$, aiming to estimate $x_1+x_2$.
However, this introduces a fractional error, which penalizes the achievable rate in the denominator ($P(a_1\Tilde{h}_2-a_2\Tilde{h}_1)^2$).
This penalty becomes more severe as $P$ increases, limiting the scheme’s scalability.
}
\end{remark}
\begin{remark}
To achieve the maximum rate in (\ref{theorem:Fading}), the user may choose $\cS_1$, $\cS_2$, and the coefficient vector $\b{a}$ to maximize it.
Specifically, this leads to the following global optimization problem:
\small
\begin{equation}\label{equ-retrieval rate for the fading AWGN MAC maximum}
\max_{\substack{\cS_1,\cS_2,\b{a}\\ a_j\neq0}}\left\{\frac{1}{2}\log^+{\left(\frac{1+P\left(\left(\sum\limits_{k\in\cS_1}h_k\right)^2+\left(\sum\limits_{k\in\cS_2}h_k\right)^2\right)}{\sqn{a}+P\left(a_1\sum\limits_{k\in\cS_2}h_k-a_2\sum\limits_{k\in\cS_1}h_k\right)^2}\right)}\right\}.
\end{equation}
\normalsize

The optimization problem involves determining the optimal partition and optimizing over a discrete space to identify the best vector $\b{a}$.
Even for a fixed $\b{a}$, this problem is analogous to the subset sum problem (or partition problem), which is NP-complete \cite{cormen2009introduction}.
However, to maximize the rate in Theorem \ref{theorem:JointFadingNew}, the user only needs to find the best partition without optimizing over a discrete space.
\end{remark}

%% file: SPIR.tex
\section{Symmetric PIR over a Block-Fading AWGN-MAC} \label{sec:symm}
The SPIR problem, a variation of the PIR problem, has attracted significant interest within the Information Theory community.
SPIR not only protects the index of the desired message from the database, ensuring user-privacy, but it also prevents the user from accessing information beyond their desired message, thereby enforcing DB-privacy. 
This implies that while the user's information is kept confidential, the user remains ignorant of the other messages in the database. 
In our context, this is mathematically expressed as \eqref{eq:DB-privacy}.

At first glance, our proposed PIR scheme might appear to satisfy the DB-privacy constraint.
This perception arises from the added responses, which seem to cancel each other out, leaving only the desired message (i.e., a 'finite field intuition' might suggest that other messages are not included in the channel output at all).
However, this is not entirely accurate.
In reality, $\mathbf{y}$ could correspond to any lattice point associated with the relevant coset of $W_\theta$, allowing the client to gain insights into the other messages.
This occurs because, while the sum of codewords before transmission is performed over the lattice, the sum 'in the air' takes place over the reals.
Consequently, the user can exploit this information and design queries to infer other messages.
Moreover, a malicious user can craft queries based on previous responses to gather information about undesired messages. Alternatively, the user could generate specific queries, permitted by the proposed scheme, to obtain a particular linear combination and leverage this to deduce undesired messages.
To illustrate this issue, consider the following example: for simplicity, assume $N=M=2$ and no dither is used (equivalent to defining the dithers as the zero vector).
We define the codebook $\cC=\{\Lambda_f \cap \cV_c\}$ using a one-dimension Nested-Lattice where $\Lambda_f=\mathbb{Z}$, $\Lambda_c=5\mathbb{Z}$,
i.e., $\Lambda_c\subseteq\Lambda_f$ (Figure \ref{fig:NestedLattice}). 
Without loss of generality, we focus on the first repetition where we assume  $\phi(W_{1}^{(1)})=1$, and $\phi(W_{2}^{(1)})=2$.
Suppose the index of the desired message is $i=1$. 
The user then generates the following queries: $Q_1(1)=[1\ 1]$ and $Q_2(1)=[0\ -1]$.
The databases form their answers $\mathbf{A}_{1}=1+2\ \text{mod} \ \Lambda_c = -2$, $\mathbf{A}_{2}=-2\ \text{mod} \ \Lambda_c = -2$.
Consequently, the user receives $\mathbf{y}=\mathbf{A}_{1}+\mathbf{A}_{2}=-4$ and is able to reliably decode $\phi(\b{W}_{1}^{(1)})$.
However, $\mathbf{y}$ can take this value only if $\mathbf{A}_{1}\ \text{mod} \ \Lambda_c=\mathbf{A}_{2}\ \text{mod} \ \Lambda_c=-2$.
Additionally, the user knows that $\mathbf{A}_{2}=-\phi(\b{W}_{2}^{(1)})\ \text{mod} \ \Lambda_c$
which implies $\phi(\b{W}_{2}^{(1)})=2$. 
Thus, the user infers both messages, indicating that the other message was leaked, meaning the suggested scheme does not satisfy the DB-privacy constraint \eqref{eq:DB-privacy}.

Adding the DB-privacy constraint \eqref{eq:DB-privacy},
intuitively necessitates that none of the database responses convey any information about the undesired message.
Fascinatingly, by making a minor modification to our PIR scheme, we demonstrate that SPIR can be achieved at the same rate as the PIR rate in Theorem \ref{theorem:JointFadingNew}, albeit with the added requirement of employing common randomness.

Let $\mathbf{S}$ be a random variable uniformly distributed over the codebook $\cC$. 
We utilize $\mathbf{S}$ as a common random codeword from the lattice codebook.
It is assumed that $\mathbf{S}$ is known only to the databases and is independent of the channel or the query. 
The following property is essential for the SPIR scheme.
\begin{lemma}\label{lemma:discret_crypto}
    For any random variable $\boldsymbol{\lambda} \in \cC=\{\Lambda_f\cap \cV_c\}$, statistically independent of $\mathbf{S}$, the sum $\mathbf{Y}=\left[\boldsymbol{\lambda}+\mathbf{S}\right] \text{mod} \ \Lambda_c$ is uniformly distributed over $\cC$ and statistically independent of $\boldsymbol{\lambda} $.
\end{lemma}
\noindent Refer to Appendix~\ref {Appendix:Crypto} for detailed proof.

Utilizing this lemma, we now proceed to prove Theorem \ref{theorem:SPIR}.
\begin{IEEEproof}[Proof of Theorem \ref{theorem:SPIR}]
Assume the user aims to retrieve message $W_i$ privately while the databases seek to prevent the user from accessing any information about the messages beyond the chosen one.
We use the same scheme as in Theorem \ref{theorem:JointFadingNew} with only one change: the databases are required to have a shared common random variable $\mathbf{S}\in\cC$ in which will be used to form their answers.
Notice that the databases must use a different common randomness for each iteration.

The databases form their answers according to the received queries in the same way as in the proof of Theorem \ref{theorem:JointFadingNew}, and add or subtract the common random variable to each answer according to their group:
\textcolor{black}{
\begin{equation} \label{eq:fading_answers_SPIR}
        \begin{split}
            \mathbf{x}_{1}^s&=[\boldsymbol{\lambda}_{1}-\mathbf{d}_1+\mathbf{S}]\text{mod} \ \Lambda_c,  \\
            \mathbf{x}_{2}^s&=\frac{\Tilde{h}_1}{\Tilde{h}_2}\mathbf{x'}_2=\frac{\Tilde{h}_1}{\Tilde{h}_2}[\boldsymbol{\lambda}_{2}-\mathbf{d}_{2}-\mathbf{S}]\text{mod} \ \Lambda_c  \textcolor{black}{.}
        \end{split}
\end{equation}
}


\textcolor{black}{
\textit{Decoding:}} To decode $\b{v}$, the user computes the following,
$$\hat{\b{v}}=\left[\alpha\frac{1}{\Tilde{h}_1}\mathbf{y}+\mathbf{d}_{1}+\mathbf{d}_{2}\right]\text{mod} \ \Lambda_c\textcolor{black}{.}$$
The reasoning follows the same arguments as in \eqref{eq:theorem_Fading_Comput} where, ultimately, the common randomness variable in each response cancels out.
Consequently, we achieve the same rate as in Theorem \ref{theorem:JointFadingNew}.

Note that the user must repeat the scheme described above $k$ times, where $k$ is the ratio between the size of the message and the size of a symbol.
Moreover, the databases should share a different common randomness for each iteration. 
Interestingly, unlike the classical SPIR problem \cite{sun2018capacity}, $\mathbf{S}$ is independent of the number of databases $N$ but depends only on the size of a codeword.


\textcolor{black}{
Notice that the user-privacy requirement is slightly different from \eqref{eq:User-privacy}, and we have to show that knowing $\mathbf{S}$ does not harm the user privacy, namely,
}
\textcolor{black}{
\begin{equation*}
    \begin{split}
&I(\theta;Q_g(\theta),W_1^M,\mathbf{x}_{g,1}(\theta),\mathbf{x}_{g,2}(\theta),\dots,\mathbf{x}_{g,k}(\theta),\mathbf{S})
\\&
\overset{(a)}{=}
I(\theta;Q_g(\theta),W_1^M,\mathbf{A}_g(\theta),\mathbf{S})
\\&
\overset{(b)}{=}I(\theta;Q_g(\theta),W_1^M,\mathbf{S})=0\\
    \end{split}
\end{equation*}
where $\mathbf{x}_{g,1}(\theta),\mathbf{x}_{g,2}(\theta),\dots,\mathbf{x}_{g,k}(\theta)$ denotes the answers of the $k$ iterations from the same database.
(a) is since there is a one to one mapping between $\mathbf{x}_{g,1}(\theta),\mathbf{x}_{g,2}(\theta),\dots,\mathbf{x}_{g,k}(\theta)$ to the elements of $\mathbf{A}_g(\theta)$.
(b) holds because $A_g(\theta)$ is a function of $Q_g(\theta),W_1^M$.
The final step follows from the mutual independence of $\theta$, $Q_g(\theta)$, $W_1^M$ and $\mathbf{S}$.
Note that for any $g\in\{1,2\}$, the query $Q_g(\theta)$ is an $i.i.d.$ $(\frac{1}{2},\frac{1}{2})$ random vector.}

\subfile{figs/LatticeExample}

We now demonstrate that database privacy, as defined in \eqref{eq:DB-privacy} is preserved.
For simplicity and clarity of notation, the proof focuses on the first iteration, where the user retrieves $W_i^{(1)}$.
Extension to multiple iterations follows naturally, as the databases utilize different instances of common randomness in each iteration, which is crucial in ensuring privacy.
\textcolor{black}{
\begin{equation*}
\begin{split}
    &I(W^{(1)}_{\overline{i}};Q_{1:2}(i),\mathbf{d}_1,\mathbf{d}_2,\mathbf{h},\mathbf{y})
    \\&\
    \overset{(a)}{\leq}
    I(W^{(1)}_{\overline{i}};Q_{1:2}(i),\mathbf{d}_1,\mathbf{d}_2,\mathbf{h},\mathbf{x}_1^s,\mathbf{x}_2^s) 
    \\&
    \overset{(b)}{=}I(W^{(1)}_{\overline{i}};\mathbf{x}_1^s,\mathbf{x}_2^s|Q_{1:2}(i),\mathbf{d}_1,\mathbf{d}_2)\\
    &=H(\mathbf{x}_1^s,\mathbf{x}_2^s|Q_{1:2}(i),\mathbf{d}_1,\mathbf{d}_2)\\
    &\qquad -H(\mathbf{x}_1^s,\mathbf{x}_2^s|Q_{1:2}(i),\mathbf{d}_1,\mathbf{d}_2,W^{(1)}_{\overline{i}})\\
    &\leq H(\mathbf{x}_1^s,\mathbf{x}_2^s)-H(\mathbf{x}_1^s,\mathbf{x}_2^s)=0\\
\end{split}
\end{equation*}
(a) is since $\mathbf{y}$ is a noisy sum of $\mathbf{x}_1^s$ and $\mathbf{x}_2^s$.
(b) follows since $Q_{1:2}(i),\mathbf{d}_1,\mathbf{d}_2,\mathbf{h},{W_1^M}$ are mutually independent, and the last step is duo to Lemma \ref{lemma:discret_crypto}.
}
\end{IEEEproof}

%% file: figs/LatticeExample.tex
\begin{figure}
\begin{tikzpicture}[scale=0.5, transform shape]
\draw[very thick,->] (-8,0) -- (8,0); 
\foreach \x in {-8,-7,...,-1} 
{
    \draw (\x cm,4pt) -- (\x cm,-4pt) node[anchor=north]  at (\x-0.13,-4pt)  {$\x$};
    \filldraw[color=blue!60, fill=blue!40, very thick](\x,0) circle (2pt);
}
\foreach \x in {0,1,...,7} 
{
    \draw (\x cm,4pt) -- (\x cm,-4pt) node[anchor=north] {$\x$};
    \filldraw[color=blue!60, fill=blue!40, very thick](\x,0) circle (2pt);
}
 \node[draw=none] at (1cm,0.5cm) {$W_1^{(1)}$};
 \node[draw=none] at (2cm,0.5cm) {$W_2^{(1)}$};
 
\foreach \x in {-5,0,5} 
{
    \filldraw[color=red!60, fill=red!40, very thick](\x,0) circle (3pt);
}
\foreach \x in {-7.5,-2.5,...,7.5} 
{
    \draw[ultra thick] (\x cm,16pt) -- (\x cm,-16pt);
}
\draw [ultra thick,decorate,decoration = {brace,raise=25pt,amplitude=10pt,mirror}] (-2.5 cm,0) --  (2.5,0)  node[pos=0.5,below=35pt,black]{Voronoi};
\end{tikzpicture}
\caption{$\Lambda_f=\mathbb{Z}$, $\Lambda_c=5\mathbb{Z}$, i.e., $\Lambda_c\subseteq\Lambda_f$.}
\label{fig:NestedLattice}
\end{figure}
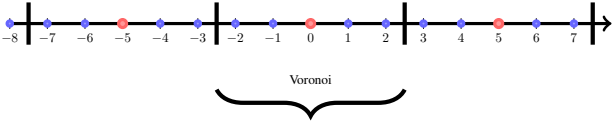

%% file: SPIR_No_Key.tex
\textcolor{black}{
\section{SPIR Without Common Randomness Among Databases}
Previous research has shown that sharing common randomness among databases is crucial for successfully achieving SPIR in classical settings, i.e, orthogonal channels and non-communicating databases.
Meaning, achieving SPIR without this common randomness is not feasible in such a setting \cite[claim 3]{gertner1998protecting}.
}

\textcolor{black}{
Interestingly, unlike the classical setting, we demonstrate that SPIR over AWGN MAC can be achieved without common randomness among the databases.}
\textcolor{black}{The significance of our results simplifies the implementation of SPIR in realistic distributed systems.
However, removing common randomness comes at the price of rate loss. }

To achieve this constructive combination of the answers without any leakage, it is crucial to eliminate the modulo-lattice operation at the servers before transmission, as these modulo operations can lead to potential information leakage (As seen in the example in Section \ref{sec:symm}). 
\textcolor{black}{Removing the modulo operation enables the avoidance of using a dither. 
Still, it requires scaling down the transmitted responses to comply with the power constraint, consequently reducing the achievable rate.}

\textcolor{black}{
For simplicity, we derive our scheme for an AWGN MAC without fading (i.e., with channel gains $h_i = 1$ for $i \in \{1,\dots,N\}$) and restrict our analysis to two databases ($N=2$). Generalizations to more than two databases and fading channels will be discussed later.
}

\subsection{Coding Scheme}
Consider a user wishing to privately retrieve the message $W_i$ while the databases seek to prevent the user from accessing any information about the messages beyond the chosen one.
\textcolor{black}{Recall the scheme described in \cite{de1975upper}, suppose that the code consists of all points of some $n$-dimensional lattice $\Lambda$ inside a hypersphere of radius $\sqrt{nP}$.
Then, a codebook $\mathcal{C}$ can be constructed whose size is equal to the ratio of the volumes of the hypersphere of radius $\sqrt{nP}$ to the hypersphere of radius $\frac{\sqrt{Mn}}{2}$ which corresponds to the effective noise level, as will be discussed later.
Thus, the second moment of a codeword $\boldsymbol{\lambda}\in\mathcal{C}$ which is uniformly distributed is $E\left[\norm{\mathbf{\boldsymbol{\lambda}}}^2\right]\leq nP$.
The codebook size is $|\mathcal{C}|=2^{nR_{code}}=(2\frac{\sqrt{P}}{\sqrt{M}})^n$.
Thus, the codebook rate is $R_{code}=\frac{1}{2}\log\left(\frac{4P}{M}\right)$.}

\textcolor{black}{
We define a mapping function $\phi(\cdot)$ that associates each vector $\mathbf{s}\in\mathbb{F}_p^l$ with a corresponding lattice codeword $\boldsymbol{\lambda}\in\mathcal{C}$. Explicitly, this mapping is given by:
\[
\mathbf{s}=(s_1,\dots,s_l)\in\mathbb{F}_p^l \mapsto \boldsymbol{\lambda}=(\lambda_1,\dots,\lambda_n)\in\mathcal{C},
\]
where the parameters $p$ and $l$ are chosen to satisfy the relation $p^l=2^{nR}$, and $R$ represents the lattice code rate.
Due to the large size of messages, they are divided into smaller chunks of length $l$ for sequential transmission, similar to the definition provided in \ref{the:th1 proof}. 
Without loss of generality, we will present the coding scheme for one such chunk.
}

\begin{IEEEproof}[\textcolor{black}{Proof of Theorem \ref{theorem:SPIRNoKey}}]

\noindent \textcolor{black}{\textit{Query:}
The user generates a random vector $\mathbf{b} \in \{-1,1\}^M$ such that each entry is either $1$ or $-1$, independently and with equal probability.
The queries constructed from the vector $\mathbf{b}$ as follows,
\begin{equation} \label{eq:SPIRnoKey_queries}
    \begin{split}
   &Q_1(i)=\mathbf{b}, \ Q_{2}(i)=-\mathbf{b} + 2b_i\mathbf{e}_i\textcolor{black}{.} \\
    \end{split}
\end{equation}
}
\textcolor{black}{Thus, $Q_1\in\{-1,1\}^M$ and $Q_2\in\{-1,1\}^M$.
Eventually, the $i$th entry of both queries will have the same sign.
Then, the user sends $Q_1(i)$ to database $1$ and $Q_2(i)$ to to database $2$. 
From the databases' perspective, each sees a uniform random vector.}  
\\
\textcolor{black}{\textit{Answers:}} 
\textcolor{black}{
Upon receiving the queries, the databases construct their responses by computing linear combinations of the messages, where the query entries determine the combining coefficients. 
That is,
\begin{equation}   \label{eq:SPIR_noKey_answers}
    \begin{split}
&\b{A}_1=\sum_{m=1}^M b_m \phi(W_m)
\\
&\b{A}_2= - \sum_{m=1, \ m\neq i}^M b_m\phi(W_m) +b_i\phi(W_i) ,
    \end{split}
\end{equation}
We note that $\b{A}_k=(a_k^{(1)},\dots,a_k^{(n)})\in \Lambda$ and not necessarily to the codebook $\mathcal{C}$.
Note that $\b{A}_1+\b{A}_{2}$ is equal to either $2\phi(W_i)$, or $-2\phi(W_i)$. 
This depends on the sign of $b_i$, which is known to the user.}

\textcolor{black}{
 To satisfy the power constraint, each database transmits a scaled form of its answer:
\begin{equation} \label{eq:NOKeytransmittion}
        \begin{split}
            \mathbf{x}_{k}&=\frac{1}{\sqrt{M}}\b{A}_k 
        \end{split}
\end{equation}
Notice that $\mathbf{x}_{1}$ and $\mathbf{x}_{2}$ are not guaranteed to be belong to $\Lambda$.
In addition
$\frac{1}{n} E[\norm{\mathbf{x}_{1}}^2] = \frac{1}{n}E[\norm{\mathbf{x}_{2}}^2] \leq P$, and that is because the messages are i.i.d and uniformly distributed.}


\textcolor{black}{
The received signal at the user antenna is then given by:
$$\mathbf{y}= \mathbf{x}_{1}+ \mathbf{x}_{2} + \mathbf{z}=\frac{2}{\sqrt{M}}\phi(W_i)+\mathbf{z}$$
}

\textcolor{black}{\textit{Decoding:}} To decode the desired lattice codeword $\boldsymbol{\phi}(W_i)$, the user first scales the received signal as:
$$\mathbf{\hat{y}}=\frac{\sqrt{M}}{2}\mathbf{y}=\phi(W_i)+\frac{\sqrt{M}}{2}\mathbf{z},$$
\textcolor{black}{Subsequently, the effective SNR of the received signal $SNR_{eff}=\frac{P}{M/4}$.}


\textcolor{black}{
According to classical results in lattice coding \cite{de1975upper}, for any rate \( R \) that is less than \( \frac{1}{2} \log(\text{SNR}) \), the probability of decoding error approaches zero as the lattice dimension \( n \) increases to infinity. 
Therefore, the achievable rate is given by,
$R_{SPIR}=\frac{1}{2} \log(SNR_{eff}) = \frac{1}{2} \log(\frac{4P}{M})$.
}

The user-privacy requirement is satisfied since, from the $m$th server point of view, the received query $Q_m(i)$ is i.i.d and uniformly distributed over $\{-1,1\}$.
As a result, each server obtains no information about the user's desired message individually.
The proof is similar to the proof of Theorem \ref{the:th1 proof}.

\textcolor{black}{
The DB-privacy constraint is satisfied by carefully constructing the server's response. 
Eventually, the user receives only the lattice codeword directly associated with the requested message $W_i$, in addition to the channel-induced noise.
Hence, no extra information about other messages is leaked.
}

\textcolor{black}{
As we can see, the rate scales well with the power $P$ and the number of databases $N$, but decreases as the number of messages $M$ increases.
Hence, while increasing transmit power improves the achievable rate, supporting more messages inherently reduces the achievable rate.
}

\textcolor{black}{Furthermore, the proposed SPIR scheme can be naturally extended to block-fading scenarios with more than two databases by adopting the grouping methodology used in the PIR scheme of Theorem~\ref{theorem:JointFadingNew}.} 
Extending the scheme to multiple databases improves the achievable rate, yielding a squared gain factor \( N^2 \) that appears within the logarithmic term of the rate expression. 
Specifically, the SPIR rate is given by
$R_{SPIR} =\frac{1}{2} \log(\frac{N^2P}{M})$.

\end{IEEEproof}

%% file: Conclusion.tex
\section{Conclusion}
\textcolor{black}{This work presents a novel PIR scheme designed for the block-fading AWGN MAC.
The proposed approach achieves higher rates than previously known results in this setting, while maintaining a bounded gap from channel capacity as the number of databases $N$ increases. 
The scheme scales favorably with both the number of databases and the transmit power $P$, making it well-suited for large-scale and high-SNR regimes.}

\textcolor{black}{Building on this foundation, we introduced two SPIR schemes that guarantee both user and database privacy.
The first scheme achieves symmetric privacy by introducing shared randomness between the databases, without compromising the achievable rate. 
The second scheme, which does not require any shared randomness, offers a more practical alternative for distributed settings. 
Although its achievable rate decreases with the number of messages $M$, it benefits from reduced implementation complexity and eliminates coordination overhead.}

\textcolor{black}{Together, these results illustrate the potential of lattice-based techniques for enabling efficient PIR over wireless MAC and provide a foundation for future work in more general and realistic communication models.}

%% file: appendices.tex
\appendices

\section{Proof of Lemma \ref{lemma:Gap}}\label{Appendix:Gap}
\begin{proof}
The proof is based on using a suboptimal, random construction for the two sets. We construct $\mathcal{S}_1$ and $\mathcal{S}_2$ to be sets of size $\floor{N/2}$, chosen uniformly from $\mathbf{h}$ without repetition, i.e., ${\mathcal{S}_1\cap \mathcal{S}_2=\emptyset}$. Assume, without loss of generality, ${\Tilde{h_1}\leq\Tilde{h_2}}$. Note that if $N$ is even, we have ${\sum_{k=1}^N |h_k|=\Tilde{h_1} + \Tilde{h_2}}$. If $N$ is odd, we have ${\sum_{k=1}^N |h_k|=\Tilde{h_1} + \Tilde{h_2}} + |h_{l}|$ for some random index $1 \leq l \leq N$, which is not in $\mathcal{S}_1$ and $\mathcal{S}_2$. Then,
\begin{align*}
    &C_{SR}^{MAC}-\max_{\substack{\mathcal{S}_1,\mathcal{S}_2 \\
    \Tilde{h}_1\leq\Tilde{h}_2}}
    \left\{\frac{1}{2}\log^{+}\left(\frac{1}{2}+\left(\sum_{k\in \mathcal{S}_1} h_k\right)^2P\right)\right\}\\
    &\leq\frac{1}{2}\log\left( 1+ P\left(\sum_{k=1}^N |h_k|\right)^2\right) -
    \frac{1}{2}\log^{+}\left(\frac{1}{2}+\Tilde{h_1}^2P\right)\\
    &\overset{(a)}{\leq} \frac{1}{2}\log\left(\frac{ 1+ P\left(\Tilde{h_1}+\Tilde{h_2} + |h_l|\right)^2}{\frac{1}{2}+\Tilde{h_1}^2P}\right)\\
    &\overset{}{=}\frac{1}{2}+\frac{1}{2}\log\left(\frac{ 1+ P\left(2\Tilde{h_2} + |h_l|\right)^2}{1+2\Tilde{h_1}^2P}\right)\\
    &\leq \frac{1}{2}+\frac{1}{2}\log\left(2\frac{ \tilde{h_2}^2 }{\Tilde{h_1}^2} +\frac{1+ 4\tilde{h_2}|h_l| P + h_l^2P}{1+2\Tilde{h_1}^2P}\right),\label{eq:break}
\end{align*}  
where $(a)$ follows since $\log^+(x)\geq\log(x)$.
Note that the elements in $\cS_1$ and $\cS_2$ are i.i.d.\ random variables with a Half-Normal distribution, mean $\sqrt{\frac{2}{\pi}}$ and variance $1-\frac{2}{\pi}$. Each set contains $\floor{N/2}$ elements. Hence, by the strong law of large numbers (SLLN), we have
$\frac{\Tilde{h_i}}{{\floor{N/2}}}\xrightarrow{N\rightarrow\infty}\sqrt{\frac{2}{\pi}}$ for $i=1,2$. Note also that 
\[
\frac{1+ 4\tilde{h_2}h_l P + h_l^2P}{1+2\Tilde{h_1}^2P} \xrightarrow{N\rightarrow\infty} 0.
\]
Since $\log(2+x) = log(2) + \frac{x}{2}+O(x^2)$, we have 
\[
C_{SR}^{MAC}-\max_{\substack{\mathcal{S}_1,\mathcal{S}_2 \\
\Tilde{h}_1\leq\Tilde{h}_2}}\left\{R_{PIR}^{eq}\right\} \leq 1 + O\left(\frac{1}{N}\right).
\]
\end{proof}



\section{Lower Bound on the Expected Achievable Rate}\label{Appendix:LowerBound}
\begin{IEEEproof}
Pick $\mathcal{S}_1$ and $\mathcal{S}_2$ as follow.
We construct $\mathcal{S}_1$ to be a set of size $\frac{N}{2}$, chosen uniformly from $\{1, ..., N\}$, and $\mathcal{S}_2$ to be $\{1, ..., N\}\setminus\mathcal{S}_1$
Given $\mathcal{S}_1$ and $\mathcal{S}_2$, compute the vector 
$\mathbf{\Tilde{h}}=(\Tilde{h}_1,\Tilde{h}_2)$.
we start with \eqref{eq:optimization},


\begin{align*}
\mathbb{E}&{}\left[R_{PIR}^{J,max}\right]
\\&
\quad=
\mathbb{E}\left[\max_{\substack{\mathcal{S}_1,\mathcal{S}_2 \\
\Tilde{h}_1\leq\Tilde{h}_2}} \left\{\frac{1}{2}\log^{+}\left(\frac{1}{2}+\Tilde{h_1}^2P\right)\right\}\right]
\\ &
\quad\overset{(a)}{\geq}
\mathbb{E}\left[
\frac{1}{2}\log\left(\frac{1}{2}+\Tilde{h^*_1}^2P\right)\right]
\\&
\quad\overset{}
{=} \mathbb{E}\left[
\frac{1}{2}\log\left(1+2\Tilde{h^*_1}^2P\right)\right]-\frac{1}{2}
\\&
\quad\overset{(b)}{=} 
\mathbb{E}\left[
\frac{1}{2}\log\left(1+2\Tilde{h^*_1}^2P\right)\Big|\left|\frac{2}{N}\Tilde{h^*_1}-\sqrt{\frac{2}{\pi}}\right|\leq \epsilon\right]
\\&
\qquad \cdot Pr\left(\left|\frac{2}{N}\Tilde{h^*_1}-\sqrt{\frac{2}{\pi}}\right|\leq \epsilon\right)
\\&
\quad +\mathbb{E}\left[
\frac{1}{2}\log\left(1+2\Tilde{h^*_1}^2P\right)\Big|\left|\frac{2}{N}\Tilde{h^*_1}-\sqrt{\frac{2}{\pi}}\right|> \epsilon\right]
\\&
\qquad \cdot Pr\left(\left|\frac{2}{N}\Tilde{h^*_1}-\sqrt{\frac{2}{\pi}}\right|> \epsilon\right)-\frac{1}{2}
\\&
\quad\overset{(c)}{\geq}
\mathbb{E}\left[
\frac{1}{2}\log\left(1+2\Tilde{h^*_1}^2P\right)\Big|\left|\frac{2}{N}\Tilde{h^*_1}-\sqrt{\frac{2}{\pi}}\right|\leq \epsilon\right]
\\&\qquad \qquad \cdot Pr\left(\left|\frac{2}{N}\Tilde{h^*_1}-\sqrt{\frac{2}{\pi}}\right|\leq
\epsilon\right)-\frac{1}{2}
\\&
\quad\overset{(d)}{\geq} 
\frac{1}{2}\log\left(1+2\frac{N^2P}{4}\left(\sqrt{\frac{2}{\pi}}-\epsilon\right)^2\right)
\\&
\qquad \qquad \cdot Pr\left(\left|\frac{2}{N}\Tilde{h^*_1}-\sqrt{\frac{2}{\pi}}\right|\leq
\epsilon\right)-\frac{1}{2} \\
\\&
\quad\overset{(e)}{\geq}
\frac{1}{2}\log\left(1+2\frac{N^2P}{4}\left(\sqrt{\frac{2}{\pi}}-\epsilon\right)^2\right)
\\&
\qquad \qquad \cdot \left(1-\frac{Var\left(\frac{2}{N}\Tilde{h^*_1}\right)}{\epsilon^2}\right)-\frac{1}{2}
\\&
\quad\overset{(f)}{\geq}
\frac{1}{2}\log\left(1+2\frac{N^2P}{4}\left(\sqrt{\frac{2}{\pi}}-\epsilon\right)^2\right)
\\&
\qquad \qquad \cdot \left(1-\frac{2(1-\frac{2}{\pi})}{N\epsilon^2}\right)-\frac{1}{2}
\\&
\quad\overset{}{=} 
\frac{1}{2}\log\left(\frac{2(2+N^2Pc)}{4}\right)-\frac{1}{2}-o(1)
\\&
\quad\overset{}{=} 
\frac{1}{2}\log\left(\frac{2+N^2Pc}{4}\right)-o(1)
\end{align*}

(a) follows from the suboptimal choice for  $\mathcal{S}_1$ and $\mathcal{S}_2$ where we denote this choise by $(\cdot)^*$.
In addition, note that $\log^+(x)\geq \log(x)$.
(b) due to the law of total probability.
(c) followed by throwing away a positive element.
(d) follows since $\left|\frac{2}{N}\Tilde{h^*_1}-\sqrt{\frac{2}{\pi}}\right|\leq \epsilon$ and thus $\Tilde{h^*_1}\geq \frac{N}{2}\left(\sqrt{\frac{2}{\pi}}-\epsilon\right)$.
we also need to set $\sqrt{\frac{2}{\pi}}-\epsilon\geq 0$ in order to get
$\Tilde{h^*_1}^2\geq \left(\frac{N}{2}\left(\sqrt{\frac{2}{\pi}}-\epsilon\right)\right)^2$.
(e) and (f) is due to the assumption of CSI at the receiver, which allows for the control of the transmitting signals' signs by sending a sign bit to each database, thereby ensuring the channel coefficients sum constructively.
Consequently, the elements in $\cS_1$ are $i.i.d.$ random variables following a Half-Normal distribution with a mean of $\sqrt{\frac{2}{\pi}}$ and a variance $1-\frac{2}{\pi}$.
Therefore, $Var(\tilde{h_1^*})=Var(\sum_{k\in\cS_1}|h_k|)=\frac{N'}{2}\left(1-\frac{2}{\pi} \right)$.
Moreover, using Chebyshev's inequality, we require that
$\sqrt{\frac{2}{N}\left(1-\frac{2}{\pi}\right)}<\epsilon<\sqrt{\frac{2}{\pi}}$.
Any $\epsilon$ outside this interval will yield meaningless results.
Thus, we set $\epsilon=0.5$ and denote 
$c=\left(\sqrt{\frac{2}{\pi}}-\frac{1}{2}\right)^2$.
\end{IEEEproof}

\section{Proof of Discrete Crypto Lemma} \label{Appendix:Crypto}
\begin{lemma}
    For any random variable $\boldsymbol{\lambda} \in\cC$, statistically independent of $\mathbf{S}$, the sum $\mathbf{Y}=\left[\boldsymbol{\lambda}+\mathbf{S}\right] \text{mod} \ \Lambda_c$ is uniform distributed over $\cC$.
\end{lemma}
We employ similar arguments for the proof as in \cite[Lemma 1]{erez2004achieving}. 
The difference between these lemmas is that $S$ is a discrete random variable, whereas the other is continuous.
\begin{IEEEproof}
    Let $\boldsymbol{\lambda}$ be any random variable taking values from $\cC$, and let the random variable $\mathbf{S}$ be uniformly distributed over $\cC$ statistically independent of $\boldsymbol{\lambda}$.
    We show that the conditional probability function $P\left(\mathbf{Y}=\mathbf{y}|\boldsymbol{\lambda}\right)$ is constant over $\mathbf{y}\in\cC$ for any $\mathbf{v}\in\cC$,
    concluding $\mathbf{Y}$ is independent of $\boldsymbol{\lambda}$:
    \begin{equation}
        \begin{split}
        &P\left(\mathbf{Y}=\mathbf{y}|\boldsymbol{\lambda}=\mathbf{v}\right)\\
        &=P\left(\left[\boldsymbol{\lambda}+\mathbf{S}\right] \text{mod} \ \Lambda_c=\left[\mathbf{y}\right] \text{mod} \ \Lambda_c|\boldsymbol{\lambda}=\mathbf{v}\right)\\
        &=P\left(\left[\mathbf{v}+\mathbf{S}\right] \text{mod} \ \Lambda_c=\left[\mathbf{y}\right] \text{mod} \ \Lambda_c|\boldsymbol{\lambda}=\mathbf{v}\right)\\
        &\overset{(a)}{=}P\left(\left[\mathbf{v}+\mathbf{S}\right] \text{mod} \ \Lambda_c=\left[\mathbf{y}\right] \text{mod} \ \Lambda_c\right)\\
        &=P\left(\left[\mathbf{S}\right] \text{mod} \ \Lambda_c=\left[\mathbf{y}-\mathbf{v}\right] \text{mod} \ \Lambda_c\right)\\
        &=P\left(\mathbf{S}=\left[\mathbf{y}-\mathbf{v}\right] \text{mod} \ \Lambda_c\right)=const\\
        \end{split}
    \end{equation}
where (a) follows since $\boldsymbol{\lambda}$ is statistically independent of $S$.
\end{IEEEproof}